\documentclass[fleqn,usenatbib]{mnras}
\usepackage{newtxtext,newtxmath}
\usepackage[T1]{fontenc}
\usepackage{ae,aecompl}
\usepackage{graphicx}	
\usepackage{amsmath}	
\usepackage{comment}
\usepackage{float}
\usepackage{color}
\usepackage{hyperref}
\usepackage{tablefootnote}
\title[The VVV Open Cluster Project]{The VVV Open Cluster Project. Near-infrared sequences of NGC\,6067, NGC\,6259, NGC\,4815, Pismis\,18, Trumpler\,23, and Trumpler\,20.}

\author[Pe\~na Ram\'irez et al.]{
K. Pe\~na Ram\'irez,$^{1}$\thanks{E-mail: karla.pena@uantof.cl} 
C. Gonz\'alez-Fern\'andez, $^{2}$
A.-N. Chen\'e, $^{3}$
and S. Ram\'irez Alegr\'ia$^{1}$
\\
$^{1}$ Centro de Astronom\'ia (CITEVA), Universidad de Antofagasta, Av. Angamos 601, Antofagasta, Chile.\\
$^{2}$ Institute of Astronomy, University of Cambridge, Madingley Road, Cambridge CB3 0HA, UK\\
$^{3}$Gemini Observatory/NSF’s NOIRLab, 670 N. A`ohoku Place, Hilo, Hawai`i, 96720, USA\\
}

\date{Accepted 2021 February 1. Received 2021 January 29; in original form 2020 October 13}
\pubyear{2020}

\begin{document}
\label{firstpage}
\pagerange{\pageref{firstpage}--\pageref{lastpage}}
\maketitle

\begin{abstract}
Open clusters are central elements of our understanding of the Galactic disk evolution, as an accurate determination of their parameters leads to an unbiased picture of our Galaxy's structure. Extending the analysis towards fainter magnitudes in cluster sequences has a significant impact on the derived fundamental parameters, such as extinction and total mass. We perform a homogeneous analysis of six open stellar clusters in the Galactic disk using kinematic and photometric information from the Gaia DR2 and VVV surveys: NGC\,6067, NGC\,6259, NGC\,4815, Pismis\,18, Trumpler\,23, and Trumpler\,20. We implement two coarse-to-fine characterization methods: first, we employ Gaussian mixture models to tag fields around each open cluster in the proper motion space, and then we apply an unsupervised machine learning method to make the membership assignment to each cluster. For the studied clusters, with ages in the $\sim$120--1900\,Myr range, we report an increase of $\sim$45\% new member candidates on average in our sample. The data-driven selection approach of cluster members makes our catalog a valuable resource for testing stellar evolutionary models and for assessing the cluster low-to-intermediate mass populations. This study is the first of a series intended to homogeneously reveal open cluster near-infrared sequences.

\end{abstract}


\section{Introduction}
Open clusters are understood as groups of coeval stars that all have the same chemical composition since, in principle, they formed in single events from a distinct molecular cloud. Besides being ideal laboratories for studying stellar structure and evolution by themselves, open clusters are spatially distributed throughout our Galaxy, making them excellent tracers of the Galactic structural, dynamical and chemical evolution. They are among the best candidates for providing precise information on both the ages and chemical compositions at various spatial positions in the disk \citep[e.g.][]{dias05,piskunov06,buckner14,jacobson16,casamiquela17}. To make significant progress, any observational endeavor in that domain need to fulfill at least two conditions: (1) accounting for a reasonably large dataset of  stellar clusters, assuring a uniform spatial distribution of stellar groups, and (2) it needs to be as homogeneous as possible, in terms of observations and processing, and to have a  consistent method to derive cluster parameters, so to minimize systematic uncertainties \citep{dias14,netopil15}. 

Large databases at optical wavelengths of open Galactic clusters in the Solar neighborhood have been available for the last two decades \citep[among others]{dias02,kharchenko13,bica19}, allowing various studies to identify new clusters and dismiss false positives \citep[e.g.][]{moitinho10,camargo16,turner17,piatti17,piskunov18}. With the arrival of the state-of-the-art astrometric information from the optical Gaia satellite releases (Gaia DR1/DR2/eDR3, \citealt{gaia16,gaia18, gaiaedr3}), it is now possible to add high-precision proper motions and parallaxes measurements to the parameter space, leading to the most complete and accurate cluster membership determination to date \citep{cantatgaudin18, castroginard18, cantatgaudin19perseus, castroginard19, castroginard20, cantatgaudin20, newcantatgaudin20, olivares19, miretroig19, galli20}. Recently, \citet{jackson20} has built up upon the kinematic space, including spectroscopic information, to assign membership probabilities using the Gaia-ESO Survey spectroscopic data on 32 open clusters.

There are a plethora of studies at near-infrared wavelengths of specific clusters \citep[e.g.][]{bonatto06,rangwal19,bisht20_4337,bisht20} as well as studies on large samples of infrared open clusters \citep{bica05,santossilva12,kharchenko16} using photometric information from the Two Micron All-Sky Survey (2MASS, \citealt{skrutskie06}). Nowadays, we can count on both a refined open cluster census via Gaia DR2 and high-quality multi-epoch near-infrared data from the Vista Variables in the V\'ia L\'actea Survey (VVV, \citealt{vvv}) and its extension, the VVVX. VVV and VVVX are ESO Public Surveys targeting the inner disc and the bulge of the Milky Way, using the VISTA telescope and its main instrument, the VISTA InfraRed CAMera (VIRCAM, \citealt{vista_2}), to
obtain multi-epoch, infrared photometry. Observations taken with VISTA are systematically reduced at CASU\footnote{http://casu.ast.cam.ac.uk/} as part of the VISTA Data Flow System (VDFS; \citealt{irwin04}). The latest data release uses v1.5 of the pipeline. More details about the performance and photometric
properties of the data can be obtained in \citet{gonzalez18}.

We present the dataset and the methodology implemented to unveil the open cluster near-infrared sequences of six open stellar clusters down to $K_s\,\sim\,15.0$\,mag: NGC\,6067, NGC\,6259, NGC\,4815, Pismis\,18, Trumpler\,23, and Trumpler\,20. We discuss how their fundamental parameters are impacted once the near-infrared sequence is unveiled.

\section{Observational data}
\subsection{Open cluster sample}
We selected a set of six clusters out of the list of clusters recovered from the literature and revisited by \citet{cantatgaudin18} and \citet{cantatgaudin20} using an unsupervised membership assignment code (see Section\,\ref{gmm_upmask}). The clusters of our sample have also been recently studied by \citet{jackson20} who added optical spectroscopic information to their procedure to determine membership. All these six clusters fall within the VVV footprint. To locate the clusters, we first have used the clusters' centers, distances, and $r50$ values from the literature. The latter is the radius (in degrees) from the center of the cluster that encompasses 50\% of the members identified by \citet{cantatgaudin20}. While it is not meant to be a physically representative description of the cluster extension, it gives an idea of the cluster's extension and density. 

We have conducted our study using five times the reported $r50$ value as the starting point for our VVV spatial tiling. None of the clusters from our sample were reported as asterisms by \citet{cantatgaudin20}. Table \ref{table:clusters} presents their identification, fundamental parameters, and the consulted references. 

\begin{table*}
\centering
\caption{Parameters of our cluster sample. Columns 2-10 contain the spatial location, proper motion, age, distance modulus, reddening, extinction, and the number of members with a  membership probability above or equal the 90\%, as defined in the literature. The number of sources identified in this work with a membership probability $\geq$90\% is presented in the last column. The number of recovered sources from the literature is indicated between parenthesis. Superscripts refer to references listed below the Table.}
\resizebox{\textwidth}{!}{%
\begin{tabular}{lcccccccccccc} 
 \hline
Name & $\alpha$ & $\delta$ & $\mu_{\alpha}*$ & $\mu_{\delta}$ & Age & (M-m)$_0$ & (M-m)$_0$ & E(B-V) & A$_\text{v}$ & Number & Number & Number\\ [0.5ex]
     &  [deg]\textsuperscript{a} &  [deg]\textsuperscript{a}  & [mas\,yr$^{-1}$]\textsuperscript{a} & [mas\,yr$^{-1}$]\textsuperscript{a} & [Myr]\textsuperscript{b} & Literature\textsuperscript{c} & Literature\textsuperscript{b} & [mag]\textsuperscript{c} & [mag]\textsuperscript{b} & p$\geq$0.9\textsuperscript{a} & p$\geq$0.9\textsuperscript{c} & p$\geq$0.9\textsuperscript{d}    \\ [0.5ex] 
 \hline\hline
 NGC\,6067    & 243.299 & $-$54.227 & $-$1.904 & $-$2.586 & 126\,$\pm$\,58   & 11.79\,$\pm$\,0.50 & 11.37\,$\pm$\,0.20 & 0.32\,$\pm$\,0.04 & 0.97\,$\pm$\,0.20 & 511 (403) & 179 (155) & 635 \\ [0.7ex] 
 NGC\,6259    & 255.195 & $-$44.678 & $-$1.015 & $-$2.892 & 269\,$\pm$\,124  & 11.91\,$\pm$\,0.53 & 11.82\,$\pm$\,0.20 & 0.69\,$\pm$\,0.06 & 1.87\,$\pm$\,0.20 & 297 (153) & 137 (71)  & 631 \\ [0.7ex]  
 NGC\,4815    & 194.499 & $-$64.960 & $-$5.754 & $-$0.964 & 371\,$\pm$\,128  & 12.87\,$\pm$\,0.85 & 12.59\,$\pm$\,0.20 & 0.67\,$\pm$\,0.11 & 1.75\,$\pm$\,0.20 & 358 (272) & 50 (23)   & 421 \\ [0.7ex] 
 Pismis\,18   & 204.227 & $-$62.091 & $-$5.658 & $-$2.286 & 575\,$\pm$\,199  & 12.43\,$\pm$\,0.69 & 12.28\,$\pm$\,0.20 & 0.69\,$\pm$\,0.05 & 1.81\,$\pm$\,0.20 & 163 (127) & 24 (18)   & 193 \\ [0.7ex] 
 Trumpler\,23 & 240.218 & $-$53.539 & $-$4.178 & $-$4.746 & 708\,$\pm$\,244  & 12.29\,$\pm$\,0.64 & 12.07\,$\pm$\,0.20 & 0.74\,$\pm$\,0.04 & 2.18\,$\pm$\,0.20 & 214 (147) & 39 (25)   & 301 \\ [0.7ex] 
 Trumpler\,20 & 189.882 & $-$60.637 & $-$7.089 & 0.180    & 1862\,$\pm$\,643 & 13.09\,$\pm$\,0.96 & 12.65\,$\pm$\,0.20 & 0.38\,$\pm$\,0.10 & 0.88\,$\pm$\,0.20 & 410 (330) & 140 (83)  & 651 \\ [1ex] 
 \hline
\multicolumn{13}{l}{\footnotesize{\textsuperscript{a}\citet{cantatgaudin20}. \textsuperscript{b} \citet{newcantatgaudin20}, $\Delta logt=0.20$ (NGC\,6067, NGC\,6259), $\Delta logt=0.15$ for the remaining clusters. \textsuperscript{c}\citet{jackson20}. \textsuperscript{d} This work. }}
\end{tabular}}
\label{table:clusters}
\end{table*}

\subsection{Multi-dimensional datasets}
We aim to identify each open cluster sequence with the most extensive dynamical range possible in the near-infrared. Although there is some variation with observational conditions, the dynamic range of the VVV survey goes from approximately (11.0, 11.0, 11.5, 12.0, 11.0) in ($Z$, $Y$, $J$, $H$, $K_s$) where detector non-linearity brings photometric errors over 0.1 to approximately (18.5, 18.0, 17.5, 16.5, 16.0\,mag) where sky noise has the same effect for a typical observing sequence. To extend our analysis to brighter sources, we combine VVV data with 2MASS photometry. We used a pre-computed nearest-neighbour, proper motion aware, cross-match between 2MASS and Gaia DR2\footnote{Using the Q3C software: \citet{Q3C}.} to retrieve  $J$, $H$, and $K_s$ magnitudes from 2MASS and the 5-parameter astrometric solution from Gaia DR2 (right ascension, declination, proper motions in right ascension and declination, and parallaxes).

We used the photometric and astrometric information from the VVV survey toward the fainter near-infrared magnitudes. A typical VISTA VVV tile was observed in between 50 to 80 epochs from 2010 to 2015. Since 2016, an extended area in both the Galactic bulge and disc is currently being surveyed as part of the VVVX survey, which should end by the time this paper is published. Both photometric and astrometric data was extracted for each source from the improved version of the VVV Infrared Astrometric Catalogue (VIRAC V2.0). We refer the reader to \citet{smith17} and \citet{clarke19} for more details, although we outline below the main steps we followed to build the catalogs.

In terms of astrometric information, the minimum spatial VISTA coverage unit (pawprint) was constructed by cross-matching the telescope pointing coordinates within a 20\,arcsec matching radius. It resulted in a sequence of images of the same on-sky region at different epochs. A VISTA tile has 2100 pawprint sets from which independent proper motions are measured. Within each of the pawprints that fulfill the selection criteria outlined by \citet{smith17}, a pool of proper motion reference sources that do not deviate significantly from the local average is extracted in an iterative process. All proper motions within a pawprint are calculated relative to this pool and corrected for any drift in $l$ and $b$ relative to Gaia DR2. As mentioned in \citet{clarke19}, the difference in drift velocity of the reference sources between pawprint sets within a VVV tile is smaller than the proper motion error obtained from a single pawprint set. To calculate final proper motions for stars observed in multiple pawprints, VIRAC 2.0 uses inverse variance weighting of the individual pawprint measurements.

\begin{figure}
\centering
\includegraphics[scale=0.3]{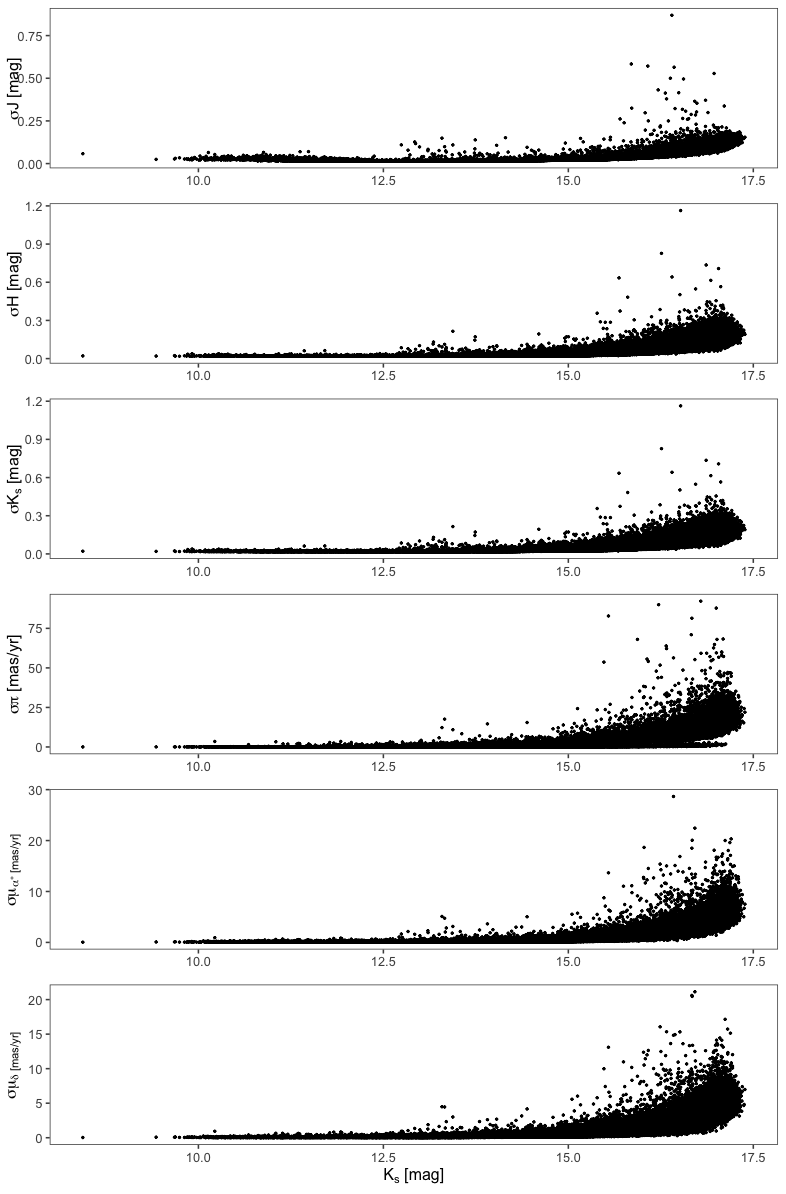}
\caption{Typical uncertainties in photometry, parallax and proper motions, as a function of $K_s$-band magnitudes for a random sample of 100000 sources in one of our initial cluster datasets.}
\label{fig:errors_sample}
\end{figure}

The final product is an astronomical dataset in eight dimensions (five astrometric parameters and three photometric ones, $J$, $H$, and $K_s$) for each cluster covering an area of five times their published \textit{r50} value. In the 2MASS/VVV magnitude overlap, sources present in both catalogues are combined with an optimal inverse variance weighting, both applied to magnitudes and proper motions/parallaxes. Since the magnitude range is broad, there is a significant difference in precision between the sources, from the brightest to the faintest. On the bright end, the nominal uncertainty reaches the Gaia precision of 0.04\,mas in parallax and 0.05\,mas yr$^{-1}$ in proper motions \citep{lindegren18}, while at $K_s\sim15.0$\,mag, the uncertainties are under $\sim$1.0\,mas (yr$^{-1}$) in parallax and proper motions (see Figure\,\ref{fig:errors_sample}). 

\section{The method}
\label{method}
\subsection{Cluster identification}\label{identify}
As the clusters of our sample are in the direction of the inner Galaxy, background/foreground contamination is problematic. In order to jumpstart our membership analysis, we construct a 2D histogram of proper motions in an circular area around the nominal cluster center of $r50$ degrees, and an equivalent histogram of a ring with the same area from $r50$ to $\sqrt{2}\times r50$. The difference of these two histograms, 
enhance the contrast of the maximum associated with the cluster. Carrying over poissonian statistics from each histogram, we located and fitted this density maximum in the residuals with a 2D gaussian. We then used this gaussian to assign a preliminar membership probability based on proper motions.

Once we had a handle on the proper motion distribution of the cluster, we mapped this into ($\alpha$, $\delta$) space, picking out sources with very low membership probability. We modelled the projected density of the remaining sources with a King profile \citep{king66} added to a constant background, assuming the coordinates of the cluster center are unknown, but with a strong prior around the nominal coordinates. From this analysis we obtained refined central coordinates and a measure of the tidal radius ($r_t$), that we used as an initial estimate of the cluster extension.

\subsection{Cluster sequence determination}\label{gmm_upmask}
One of the main challenges when one attempts to spatially characterize a stellar conglomerates is the contrast between a cluster and the field stars, which is most of the time shallow. And it is a challenge that the motions of the stars due to proper motions and parallaxes only make harder. Nevertheless, it should be possible to identify cluster members even among stars with significant astrometric uncertainties if other criteria such as photometric selections are employed.

We revisited the near-infrared cluster sequences following a methodology as presented in \citet{cantatgaudin19perseus}. For each cluster field, we first applied the Gaussian Mixture Model \citep[GMM; e.g.][]{everitt11} technique, which is based on the assumption that the stars distribution within an overdensity can be described by a superposition of multivariate gaussian distributions \citep{desouza17}. The GMM was implemented in \textsc{R} \citep{Rref} using the \textsc{mclust} library \citep{mclust16}. For each cluster, we applied the GMM technique to the overdensity detected in the proper motion space, using only the stars within the $r_t$ radius (as described in Section\,\ref{identify}) and inside 1-$\sigma$ of the distance of the cluster as estimated with Gaia. We used the parallax inverse as a proxy to the distance, since we implicitly assume that all the stars in the clusters are roughly at the same distance. We used ten multivariate gaussian components per overdensity, yet the results are insensitive to the exact number of gaussians, as long as it is large enough to disentangle the background from the members. Any gaussian component with a covariance bigger than 0.04\,mas$^2$\,yr$^{-2}$ in both $\mu_{\alpha}*$ and $\mu_{\delta}$ (corresponding to a standard deviation of $\sim$0.2\,mas\,yr$^{-1}$) and/or a mixing probability higher than $15\%$ was rejected. As expected, we were left only with gaussian components centered on proper motion values within the uncertainties of the Gaia values. The total number of sources populating the remaining gaussians ranged from 1616 for NGC\,6259 to 211 for Pismis\,18.

Once the GMM analysis was completed, we employed the Unsupervised Photometric Membership Assignment in Stellar Clusters \textsc{UPMASK} \citep{upmask14} on each cluster region. This approach has successfully been applied to Gaia DR2 data \citep{cantatgaudin18, cantatgaudin19perseus} before we applied it to our near-infrared data. The key assumptions are:
\begin{enumerate}
    \item the cluster members share common properties (i.e., they are clustered in the magnitudes and colors spaces).
    \item their spatial distribution is not compatible with a uniform one.
\end{enumerate}

The coherent motion of the conglomerates did not need to be verified at this point, since the input data was already clustered in the 2D proper motion space ($\mu_{\alpha}*$, $\mu_{\delta}$) after the fist pass using the GMM. The first step is to group stars according to their spatial and photometric distribution using the $J$, $H$, $K_s$ photometry and the spatial positions. To do so, we used the \textsc{k-means} clustering \citep{kmeans} partition algorithm with a large k to the dataset size guaranteeing at least 25 objects per group. The second step is to test whether the distribution of stars within each group is more concentrated than what is expected for a random fluctuation in a uniform distribution. In this implementation, we use the total length of a minimum spanning tree \cite[e.g.][]{mst} and we iterate 100 times per cluster field. At each iteration, the photometry data were randomly sampled from the probability distribution function of each star's positional parameters, while taking into account uncertainties for each variable. The clustering score for a given source is derived directly from the number of iteration during which it was member of a concentrated group, and can be interpreted as a membership probability. 

\begin{figure*}
\includegraphics[scale=0.1]{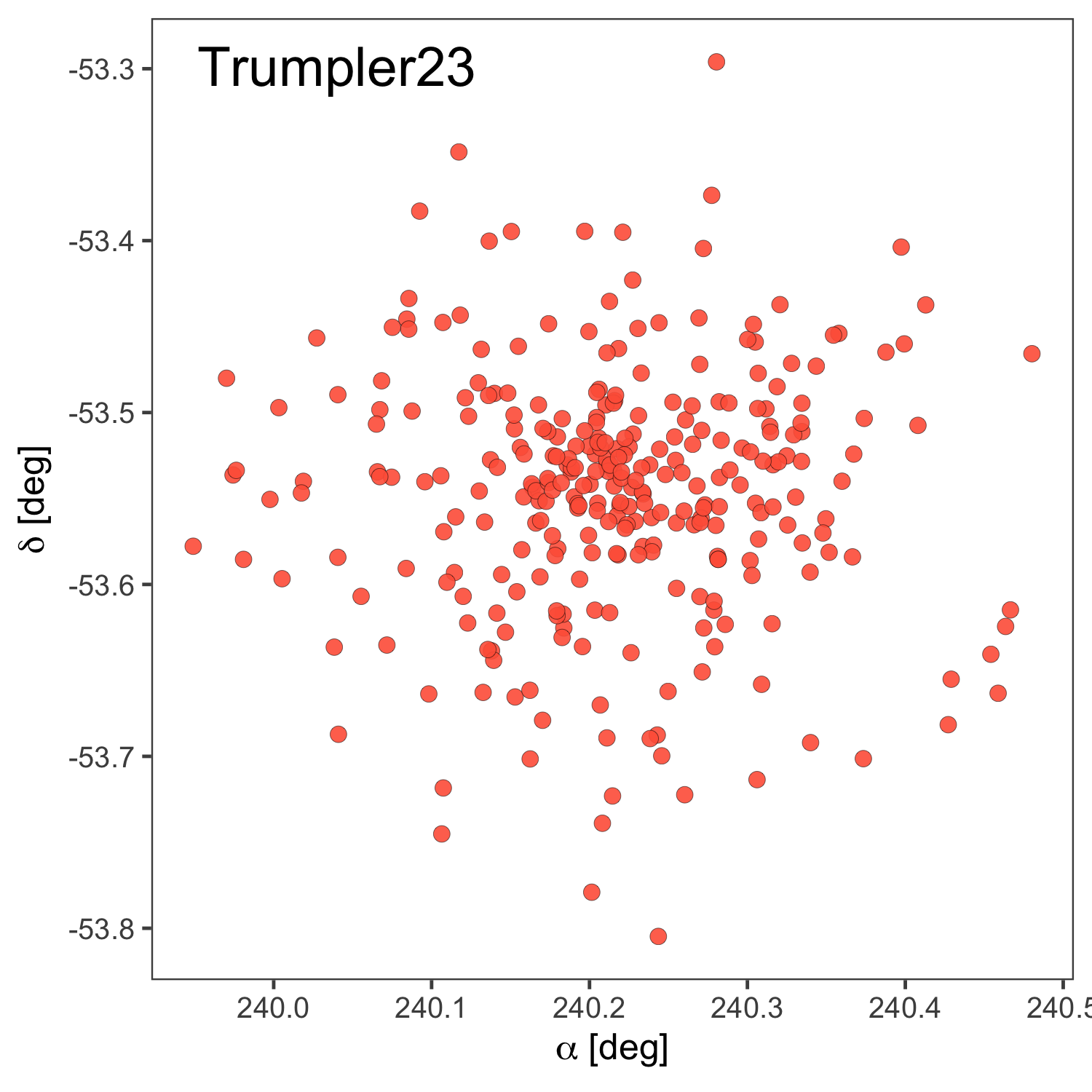}
\includegraphics[scale=0.1]{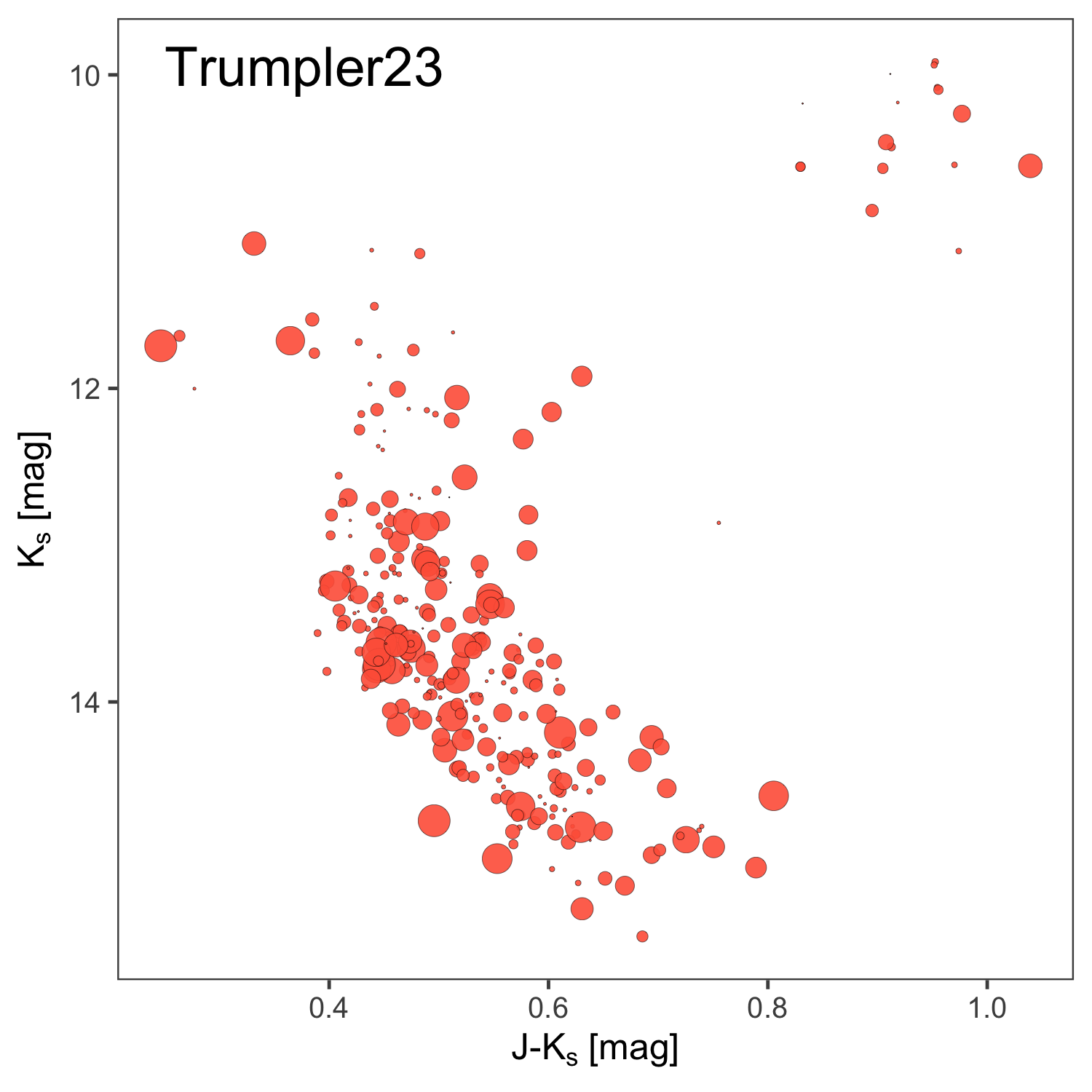}
\includegraphics[scale=0.1]{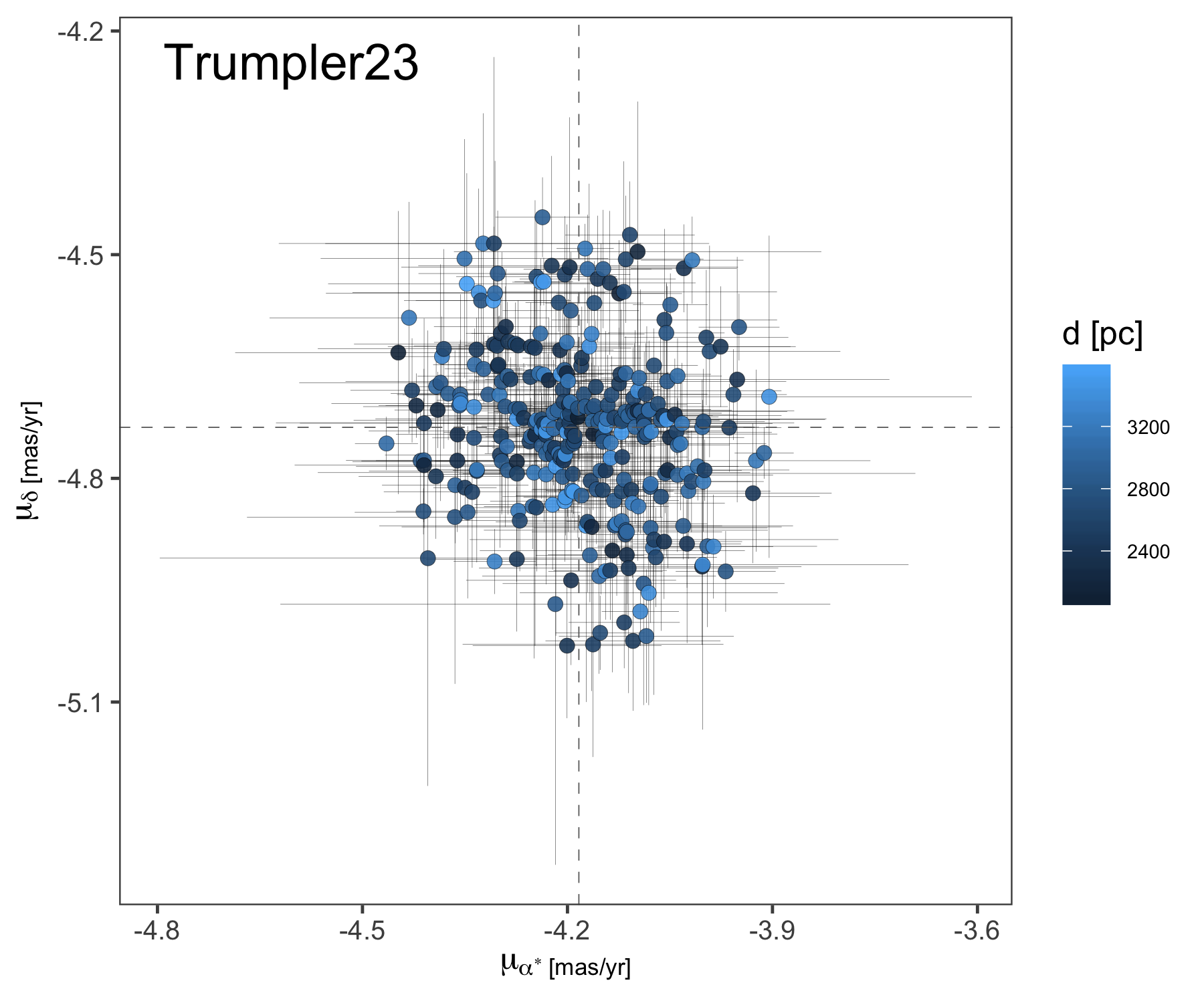}\\
\includegraphics[scale=0.1]{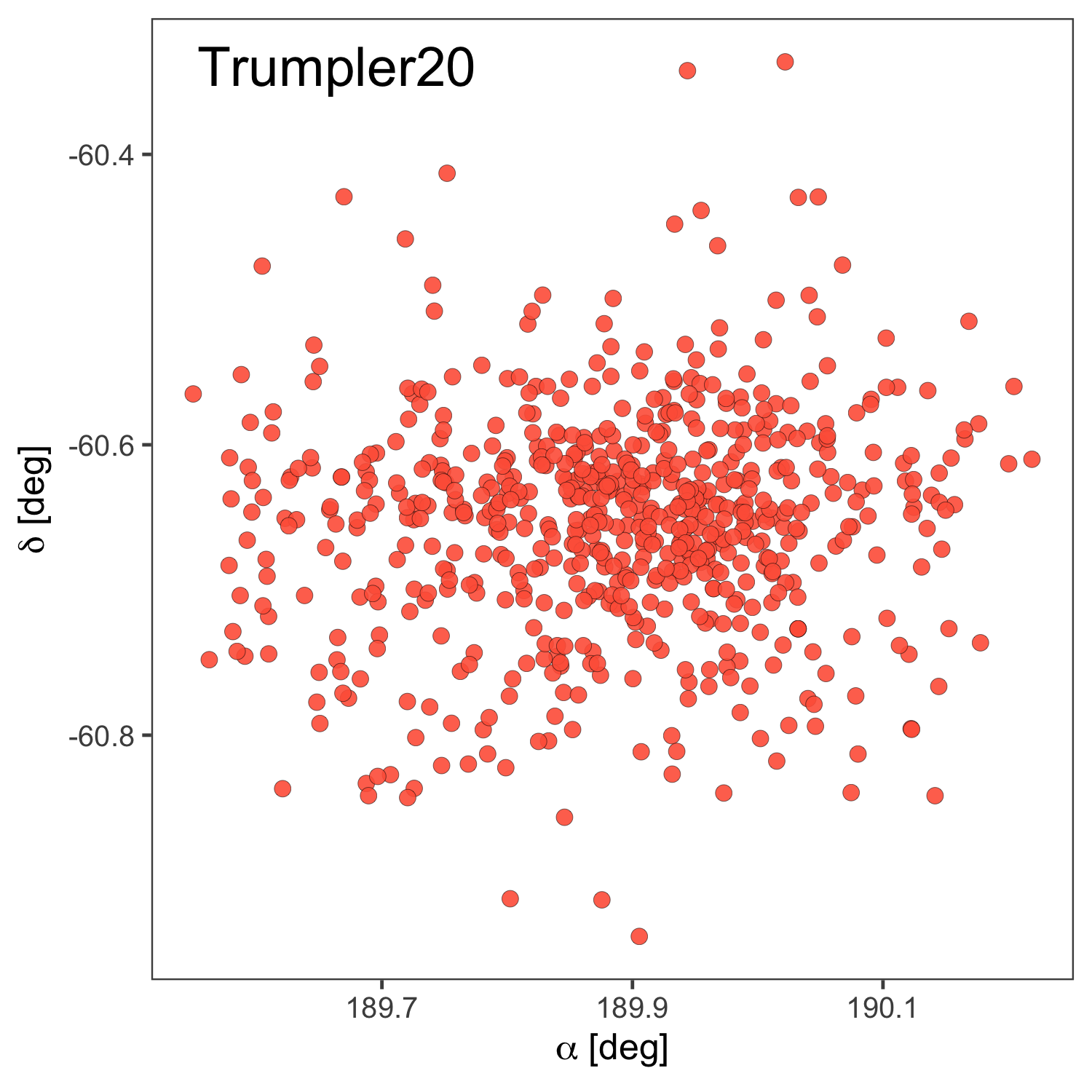}
\includegraphics[scale=0.1]{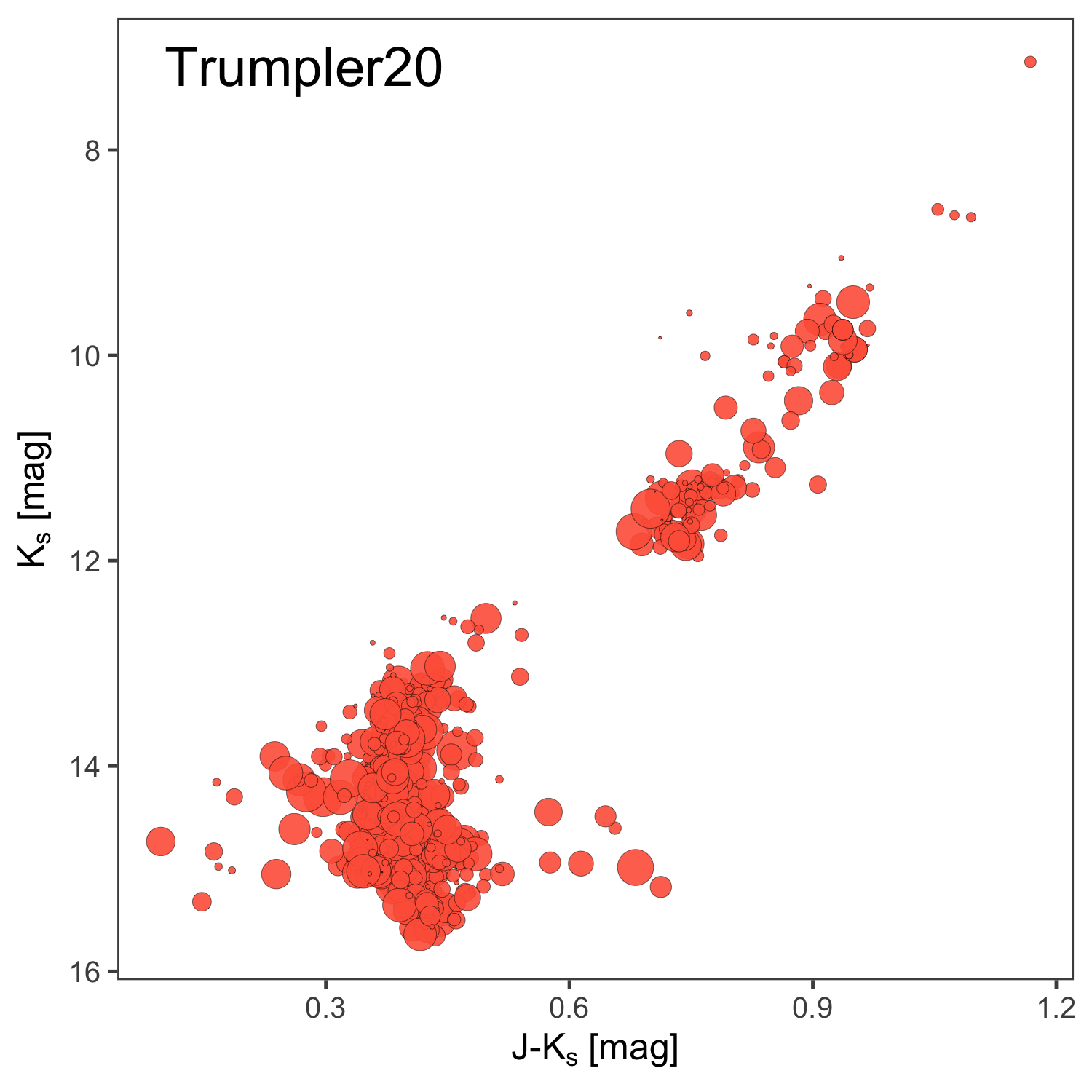}
\includegraphics[scale=0.1]{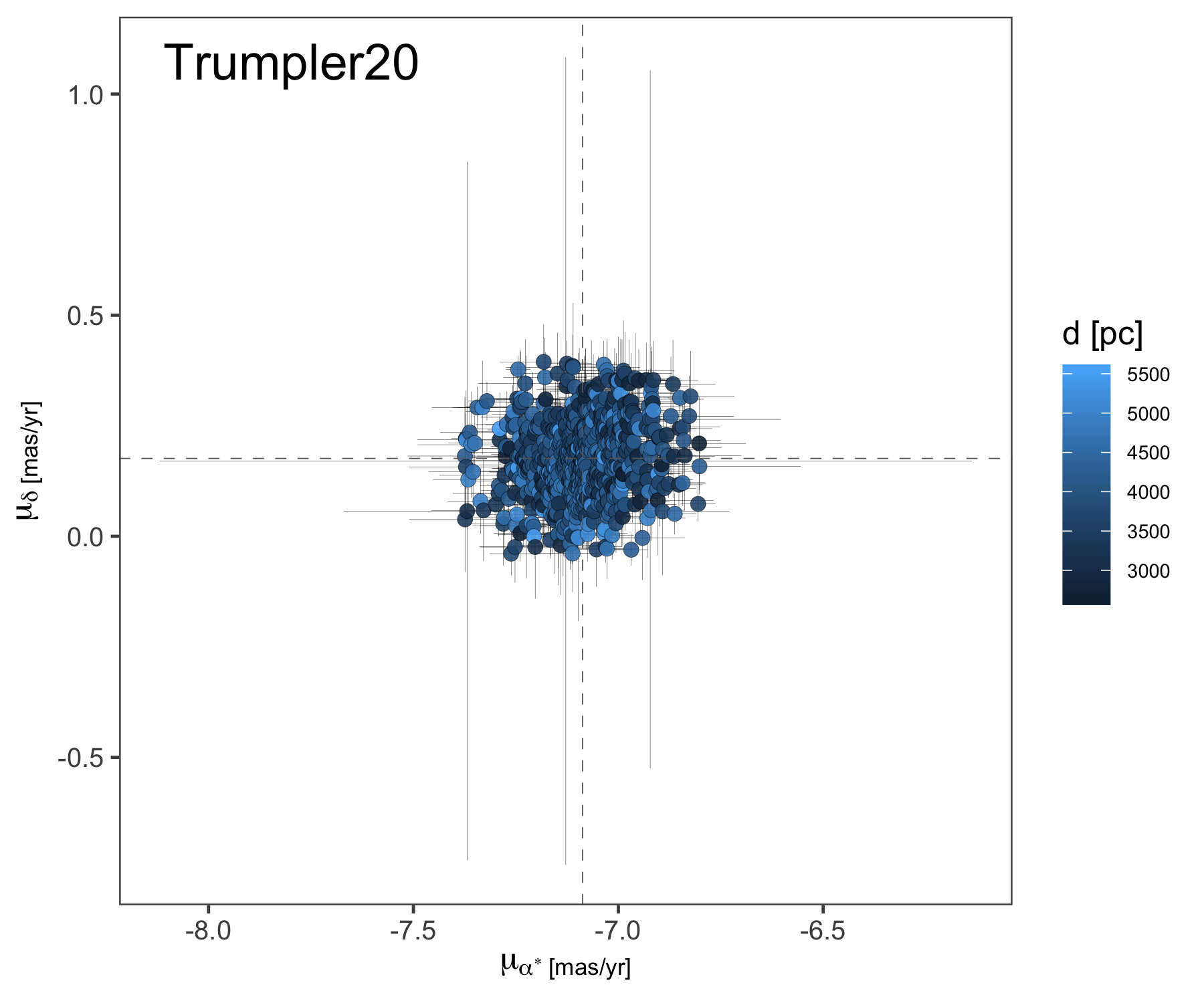}\\
\caption{\textit{Left:} Spatial distribution of the cluster members. \textit{Middle:} $K_s$ vs. $J$-$K_s$ color-magnitude diagrams of the studied open stellar clusters. The point size scales with the distance to the cluster center. \textit{Right:} Members proper motion distribution. The uncertainties are color-coded based on their distance value. Mean proper motions in right ascension and declination are plotted with dotted lines. Only sources with membership probabilities $p\geq$90\% are shown in all the diagrams. Text on the plots shows their corresponding name.
}
\label{fig:clusters}
\end{figure*}

Our clustering score is then based on the 5D ($\alpha$, $\delta$, $J$, $H$, $K_s$) information of each star and associated nominal uncertainties. Figure\,\ref{fig:clusters} shows the spatial distribution, near-infrared sequence and, proper motions for each cluster. Complementary, Figure\,\ref{fig:clusters_info} shows the clusters near-infrared sequences with the different membership ranges color-coded as well with the available literature data overplotted. The full membership list per cluster is available in electronic form (see Section \ref{catalogs}).

\begin{figure*}
\includegraphics[scale=0.1]{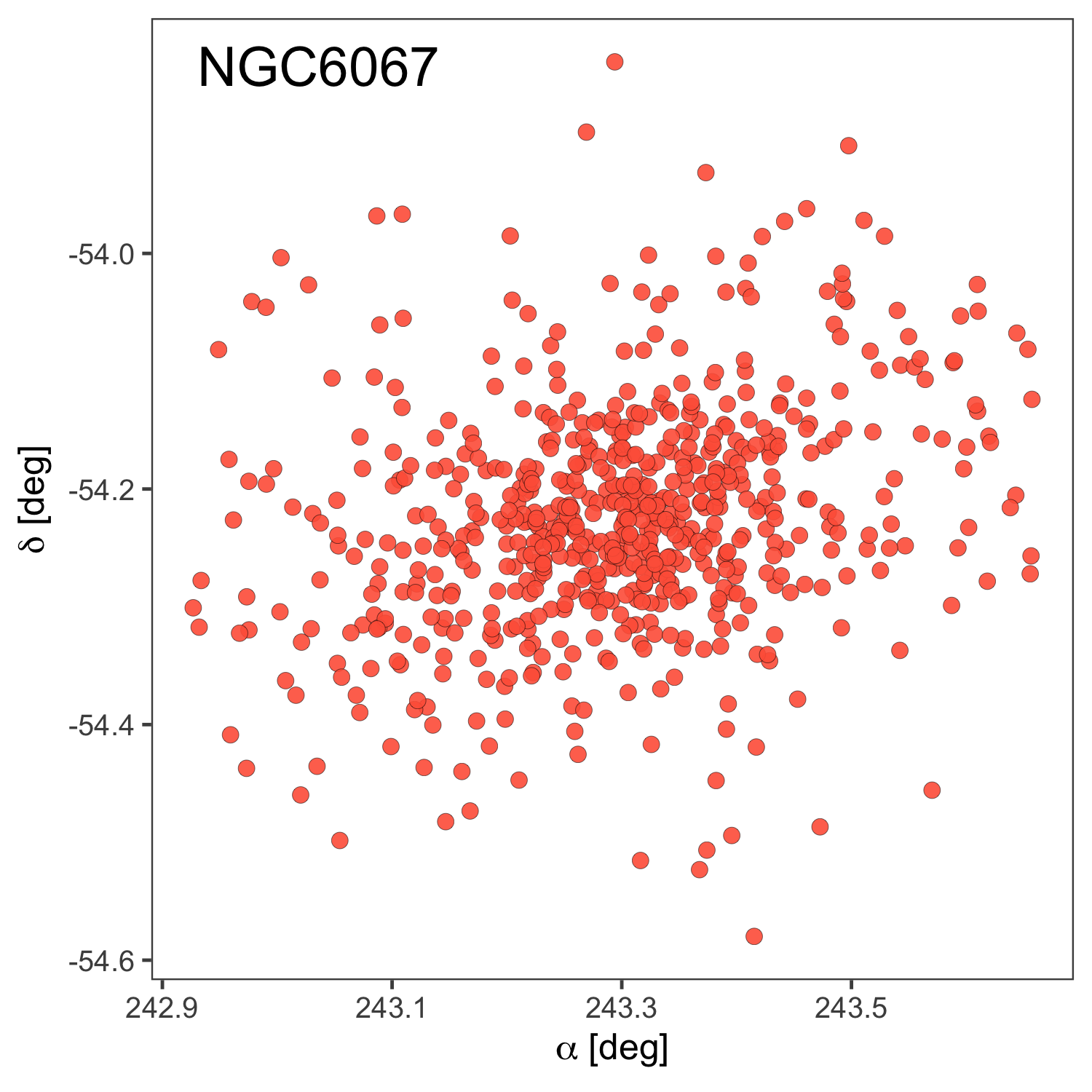}
\includegraphics[scale=0.1]{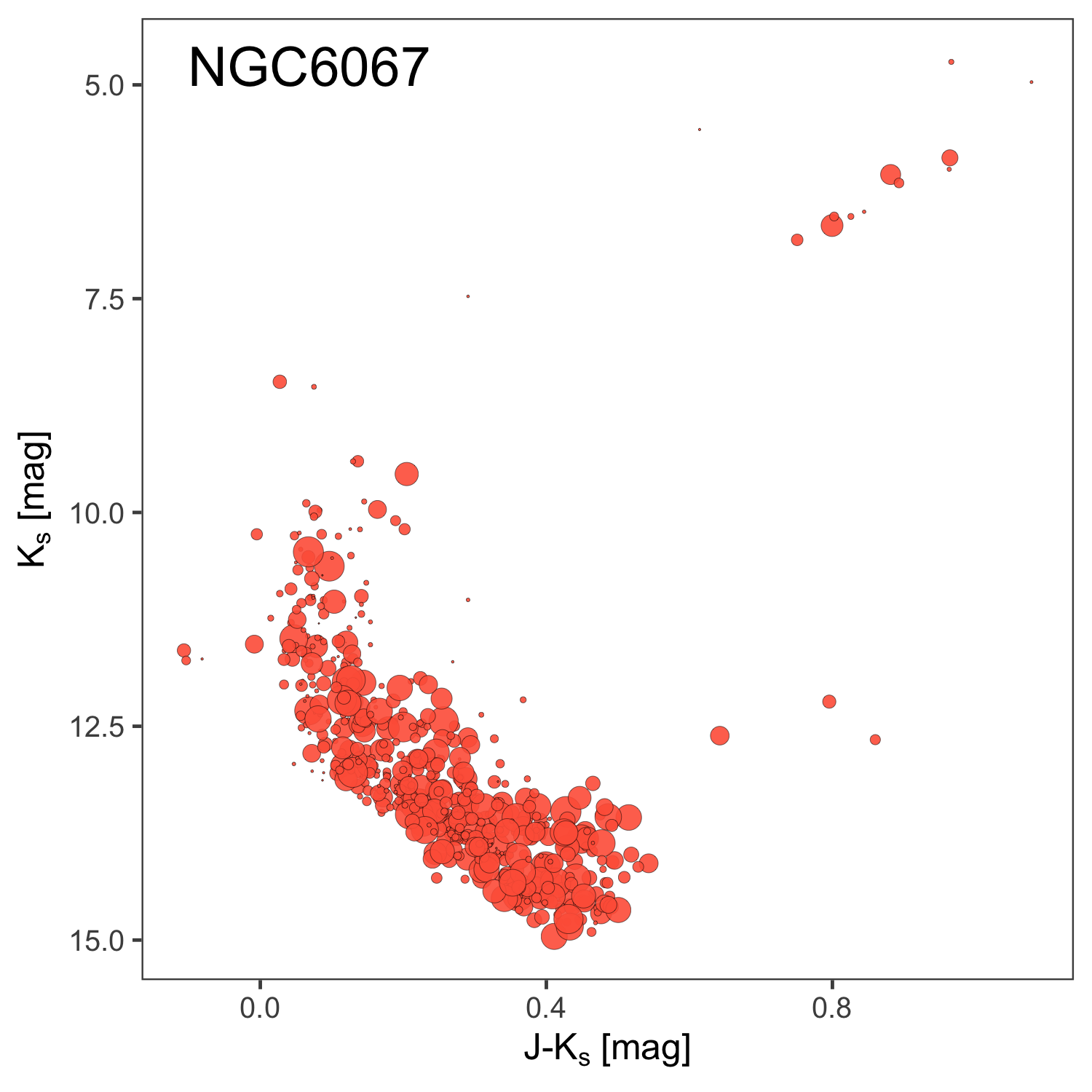}
\includegraphics[scale=0.1]{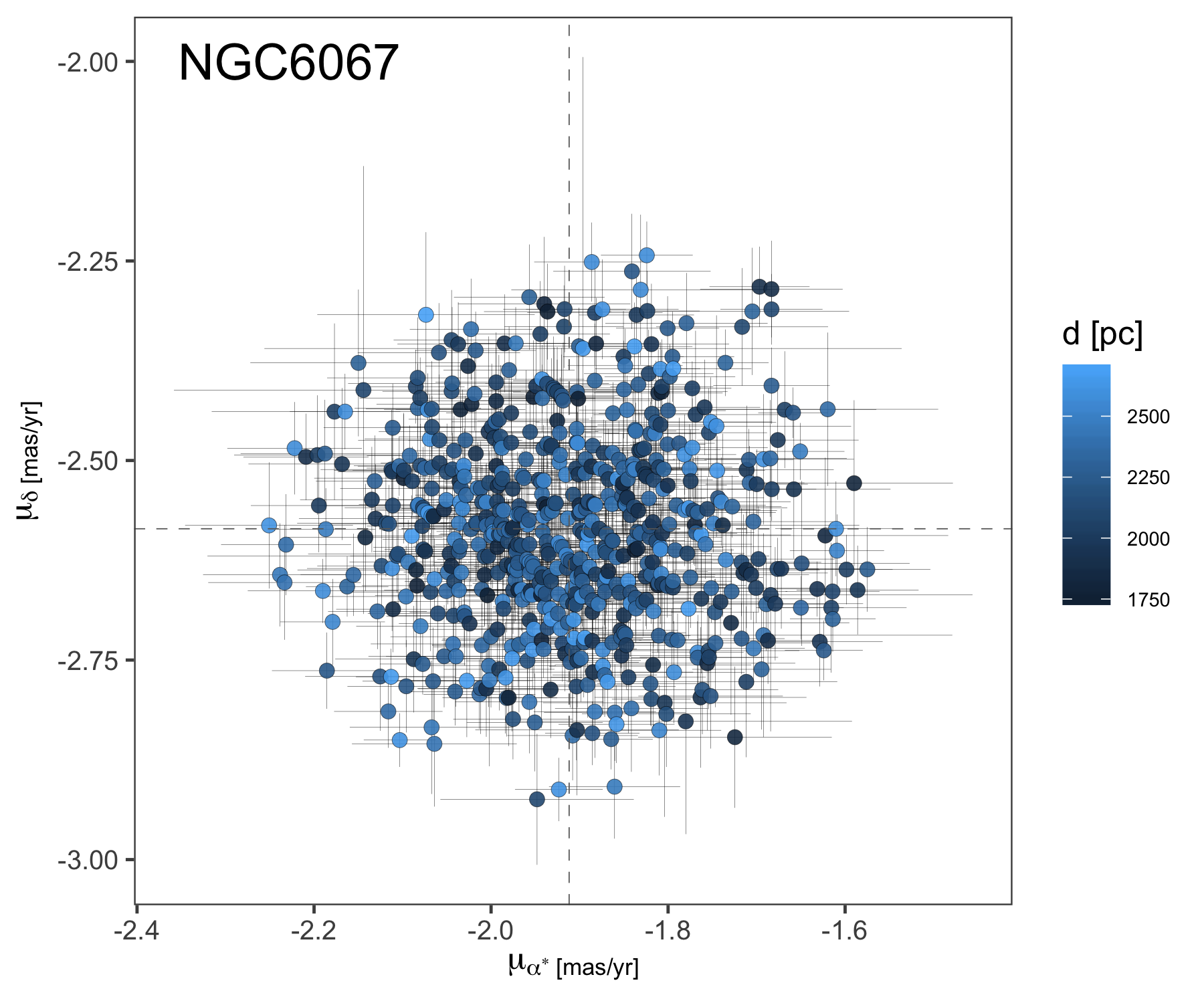}\\
\includegraphics[scale=0.1]{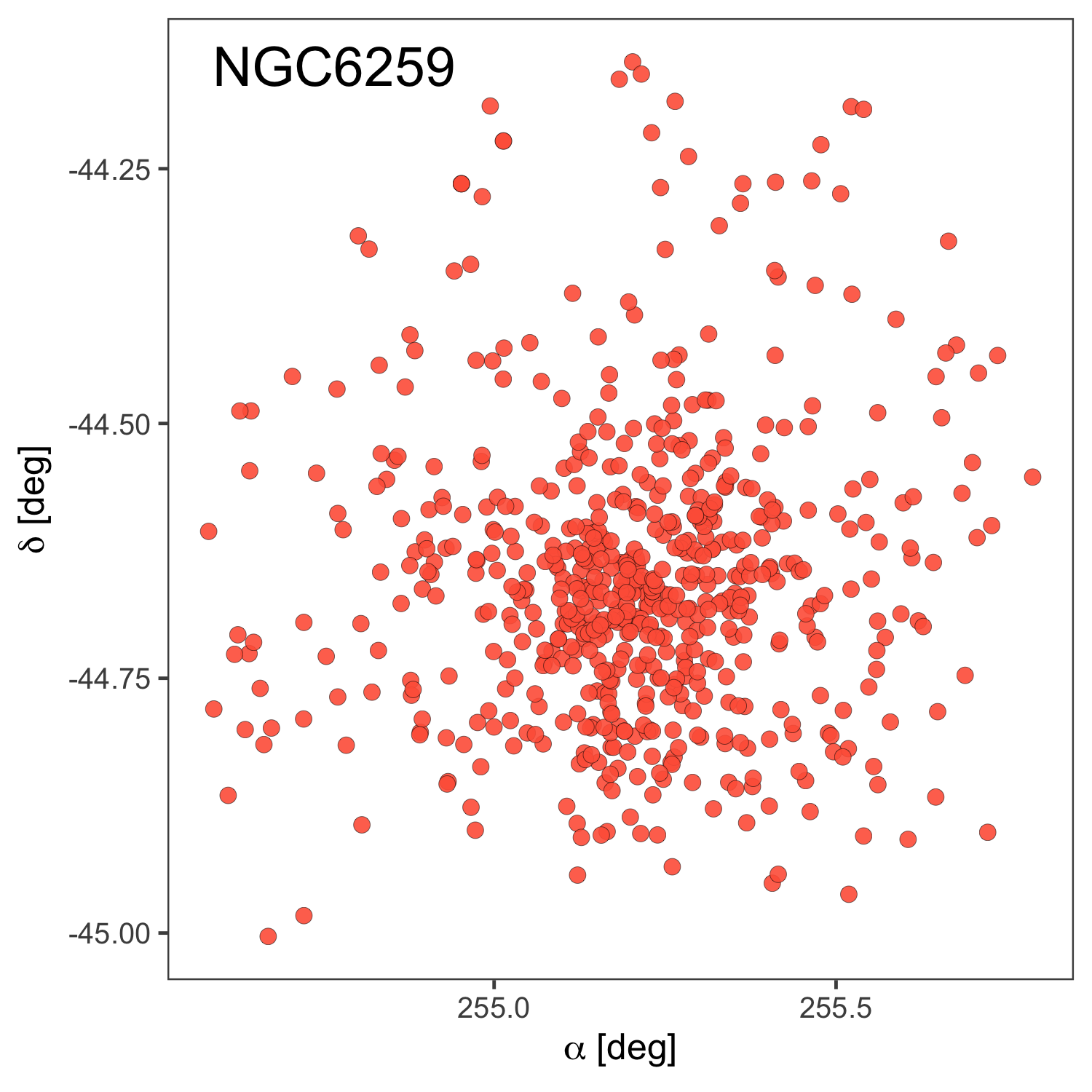}
\includegraphics[scale=0.1]{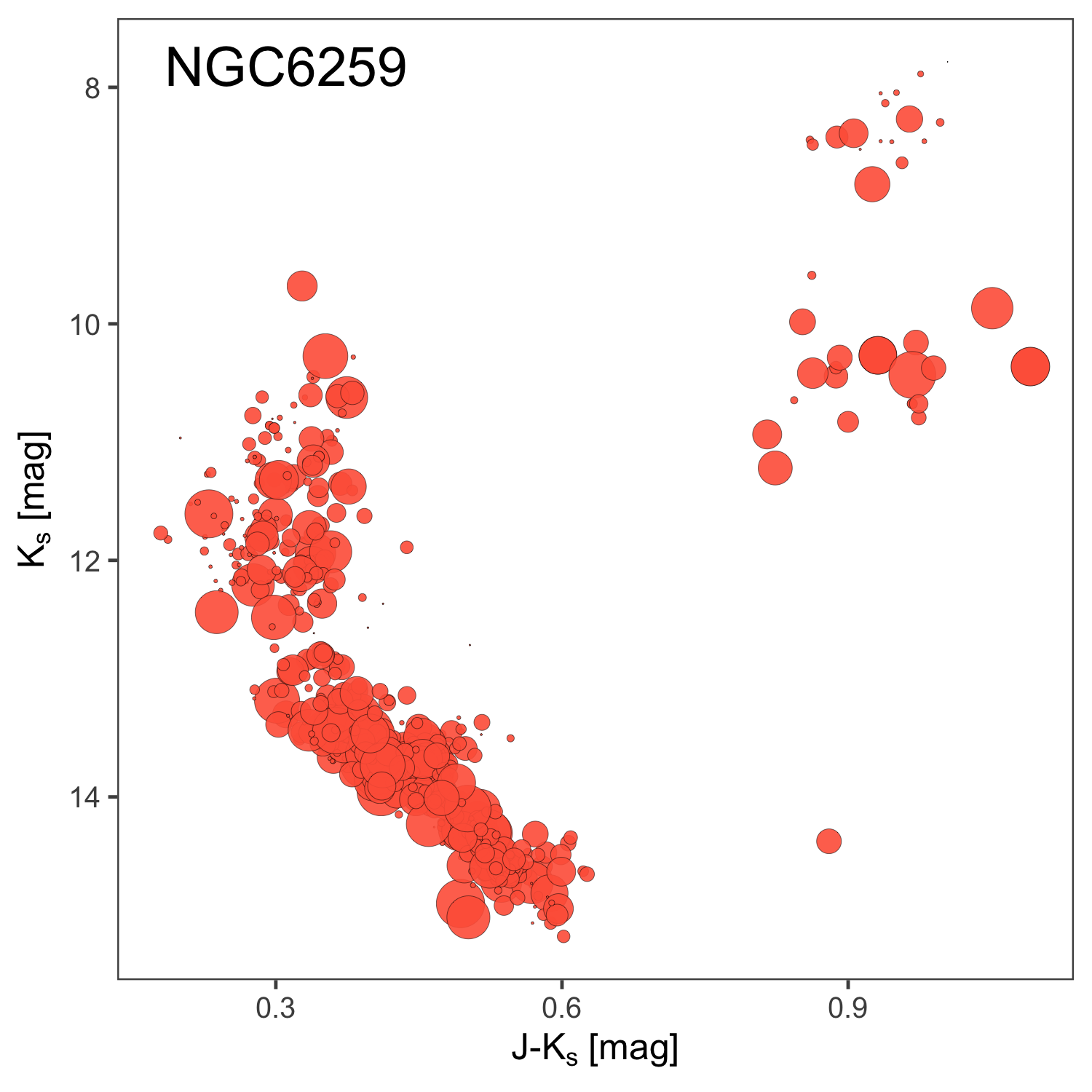}
\includegraphics[scale=0.1]{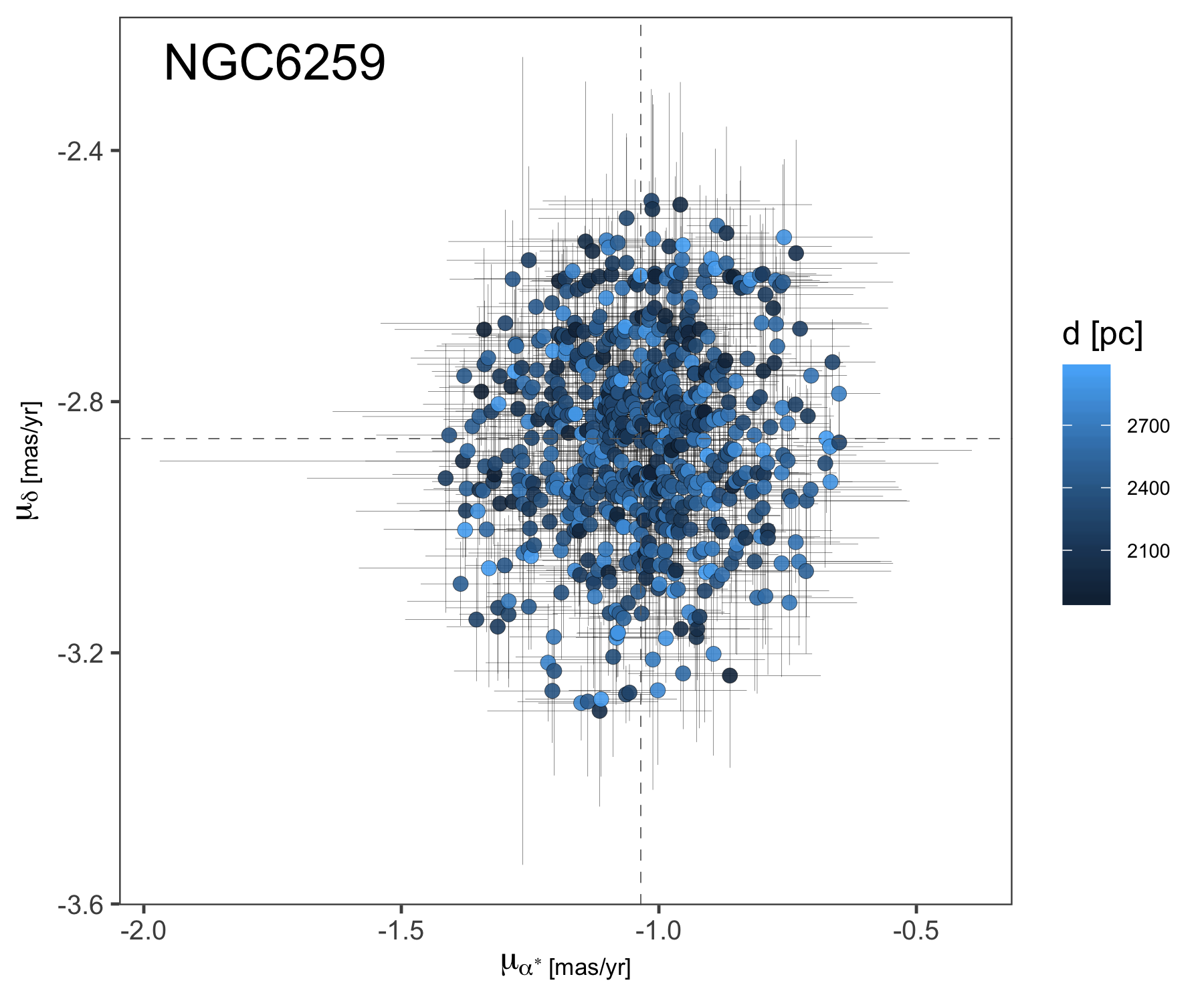}\\
\includegraphics[scale=0.1]{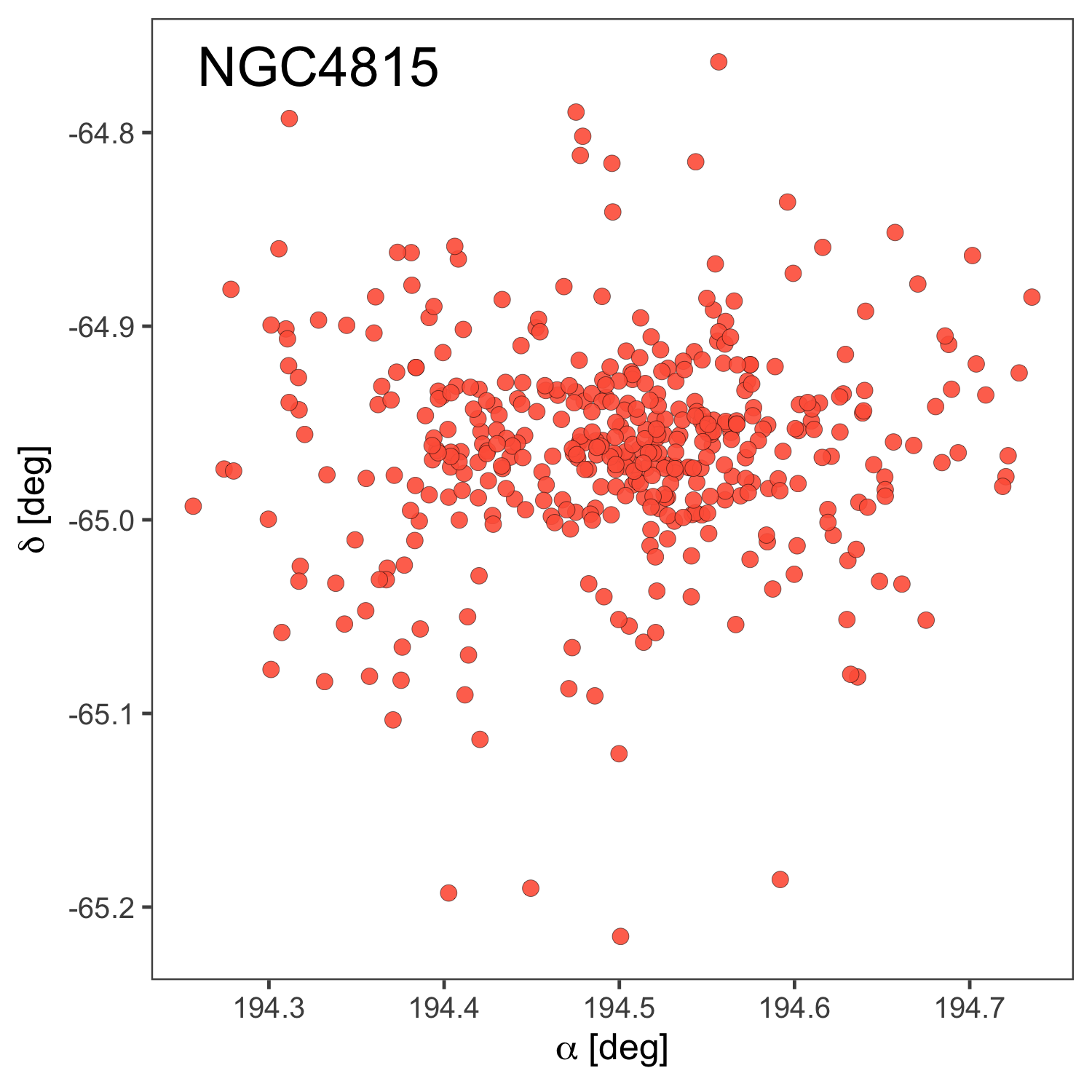}
\includegraphics[scale=0.1]{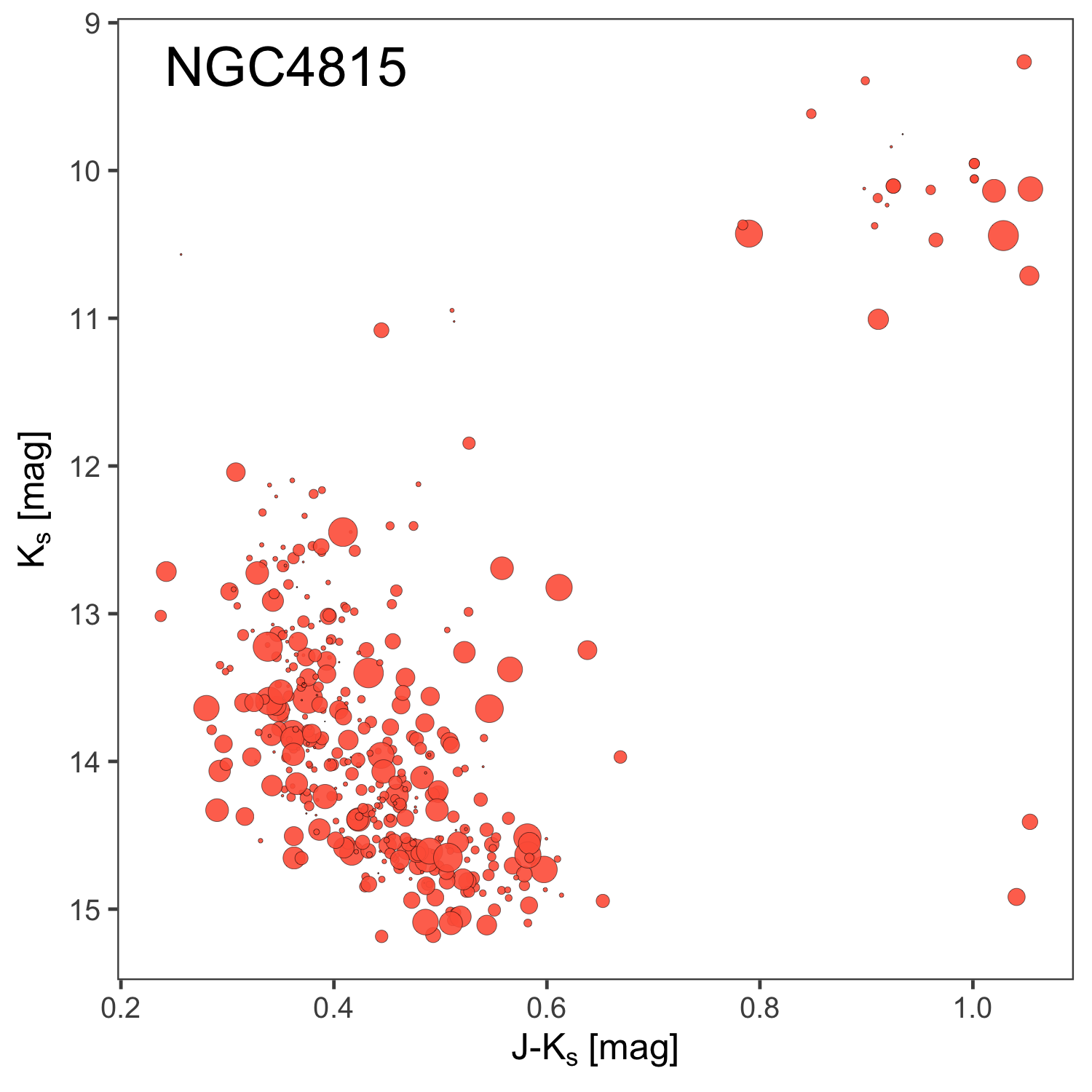}
\includegraphics[scale=0.1]{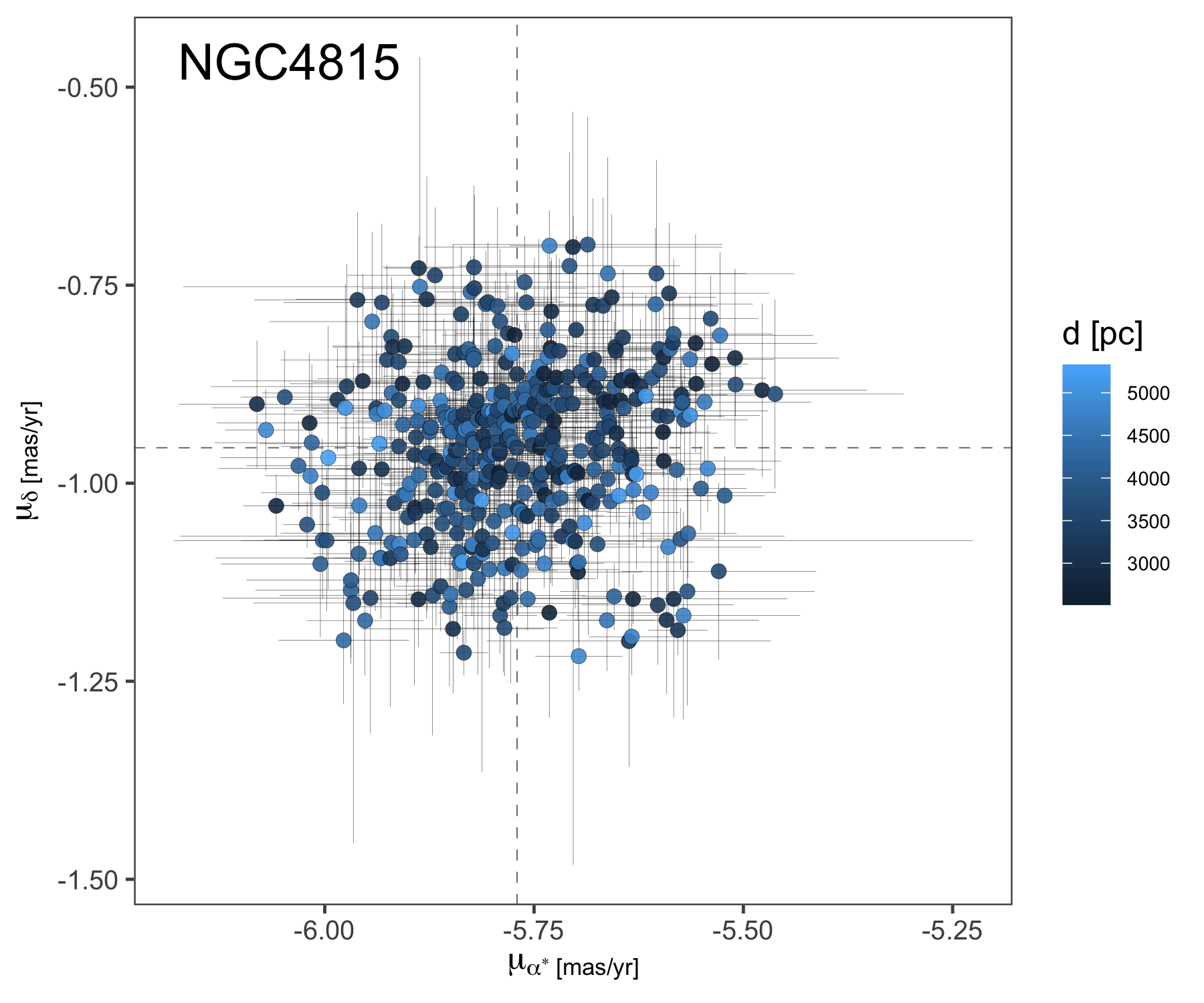}\\
\includegraphics[scale=0.1]{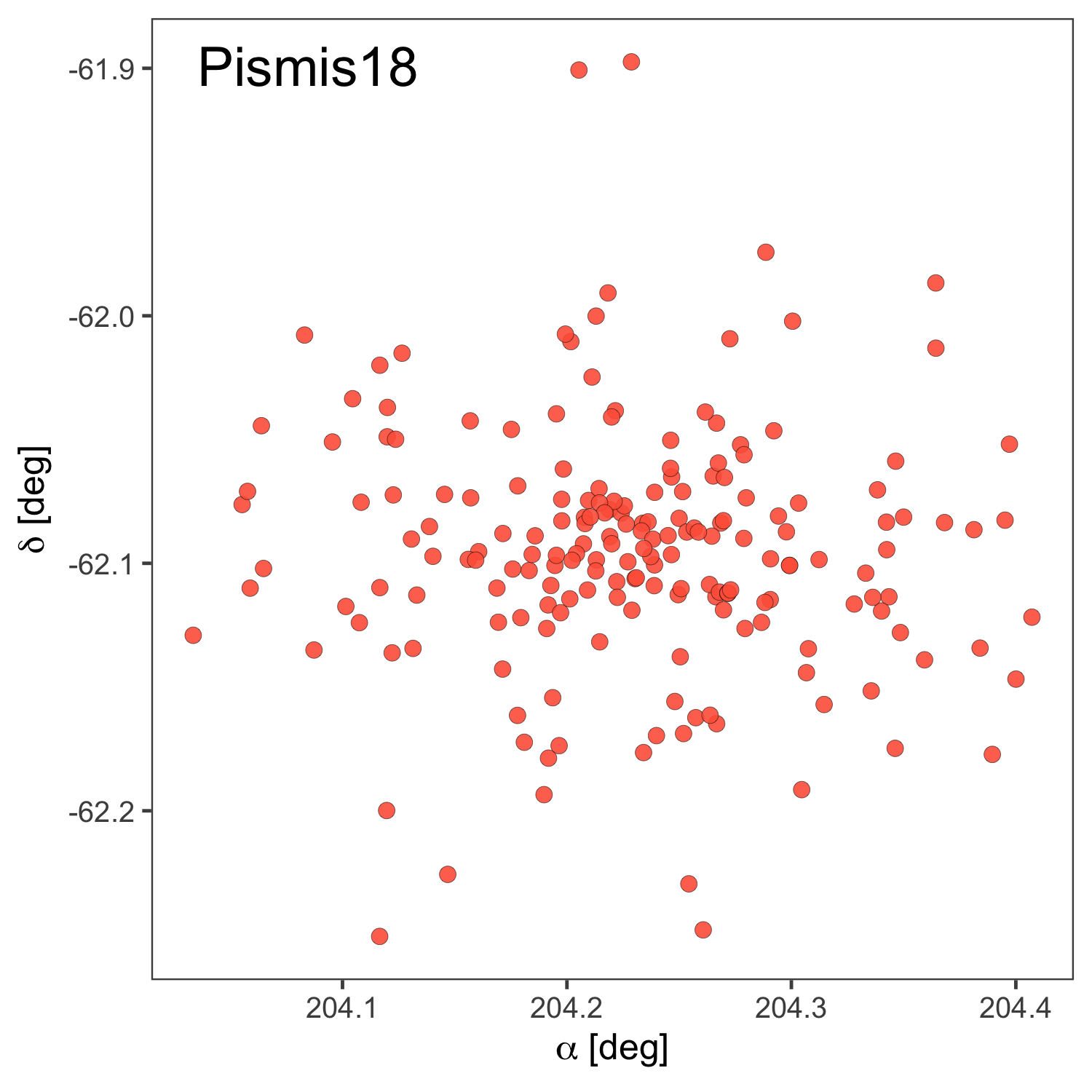}
\includegraphics[scale=0.1]{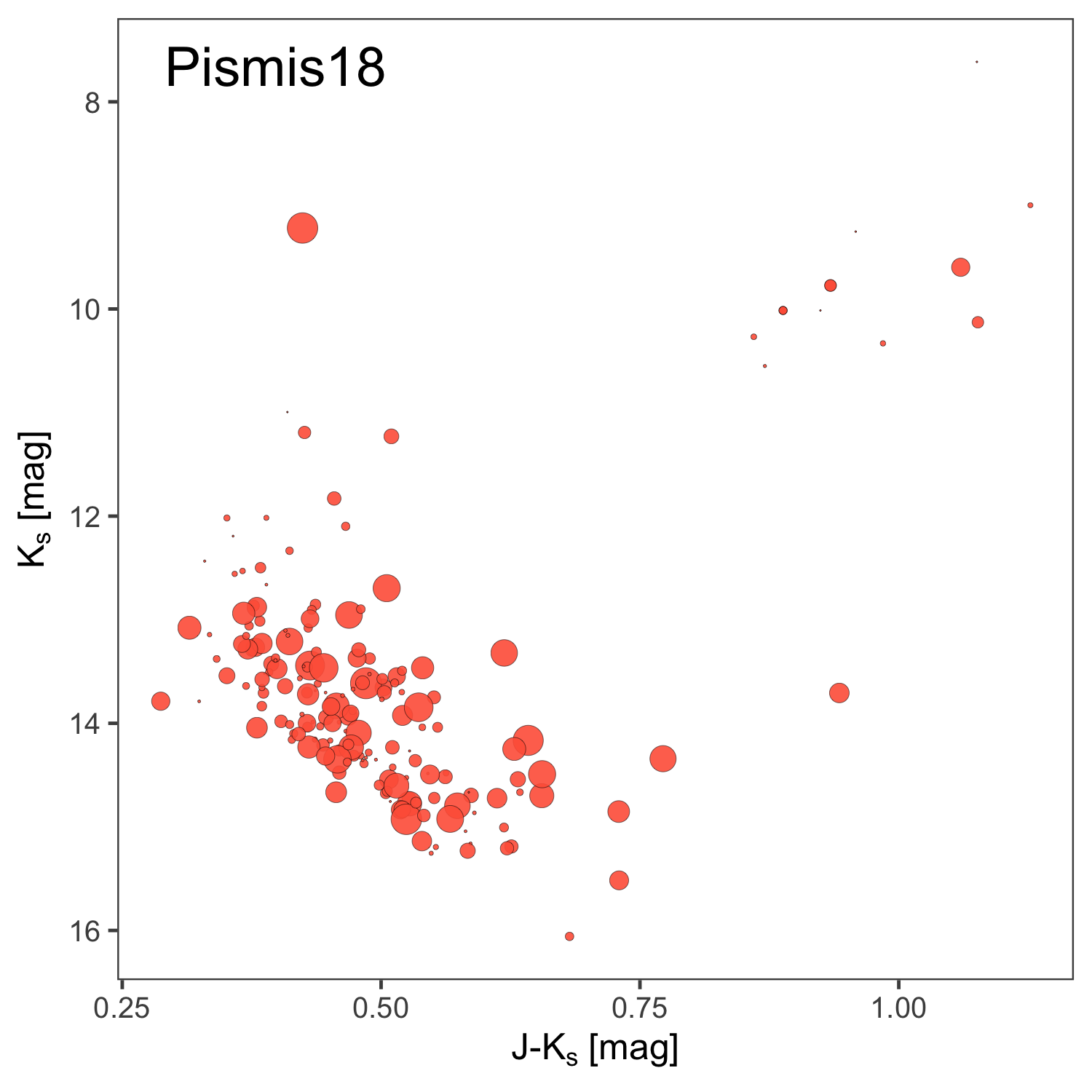}
\includegraphics[scale=0.1]{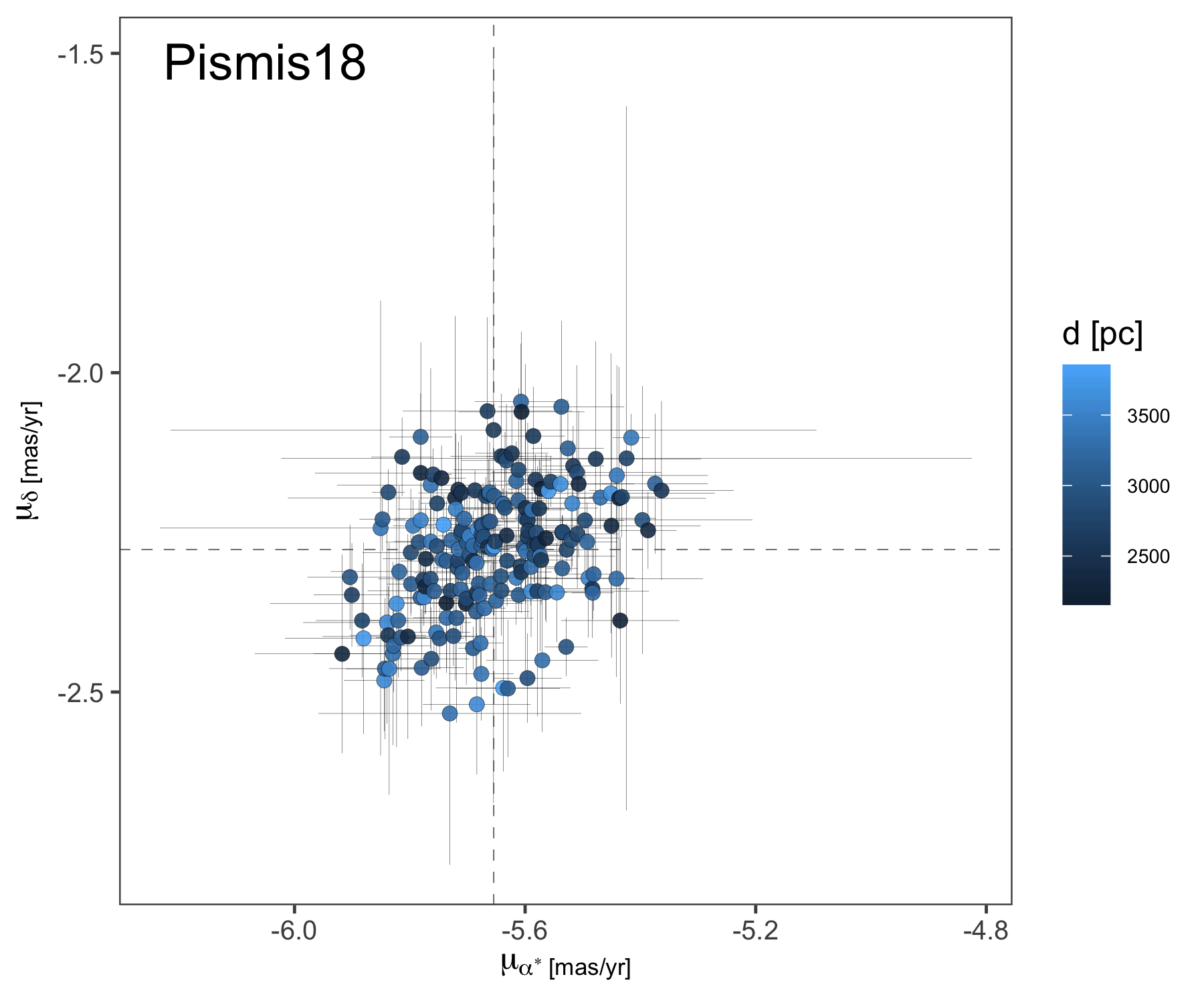}\\
\includegraphics[scale=0.1]{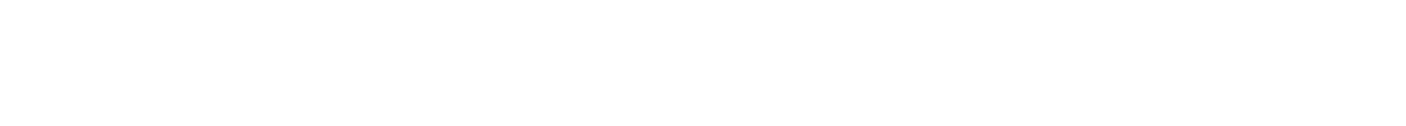}\\
\end{figure*}

\section{Discussion}
\label{sec:discussion}
High precision characterization of star clusters fundamental parameters, such as age, distance, reddening, and total mass, depends on the quality of the cluster membership determination. And the more dimensions are considered in membership determination, the better; therefore the need for datasets that are as diverse as they are accurate. 
The decontamination of background interlopers must ensure than non-additional biases are imposed, and that whatever cluster members remains is indeed robust and representative of the cluster
itself.

Using only sources with a membership probability larger or equal than 90\%, we have re-derived the median spatial position, proper motion, and distance for each cluster. We used the mean absolute deviation (from the median), or MAD values \citep{feigelson12}, to estimate the dispersion of those values. The results are gathered in the Table\,\ref{table:values}, together with the derived $r_t$ value. The offset between our recalculated central positions and those from literature range from 0.01 to 1.5\,arcmin, the larger value corresponding an offset in declination calculated for NGC\,6259. 

\begin{table*}
\centering
\caption{Center, proper motion, tidal radius, distance, reddening, extinction, and total mass values for the studied clusters.}
\begin{tabular}{lcccccccccccc} 
 \hline
Name & $\alpha$ & $\delta$ & $\mu_{\alpha}*$ & $\mu_{\delta}$ & Age\textsuperscript{a} &  $r_t$ & d & E($J-K_s$) & $A_{K_s}$ & M \\ [0.5ex]
     & [deg] & [deg] & [mas\,yr$^{-1}$] & [mas\,yr$^{-1}$] & [Myr] & [arcmin] & [pc] & [mag] & [mag] & [M$_{\odot}$]\\ [0.5ex] 
 \hline\hline
 NGC\,6067 & 243.296 & $-$54.225 & $-$1.91\,$\pm$\,0.13 & $-$2.58\,$\pm$\,0.12 & 126\,$\pm$\,58 & 23.4 & 2230\,$\pm$\,269 & 0.15 & 0.7 & 3030\,$\pm$\,40\\ [0.7ex] 
 NGC\,6259 & 255.204 & $-$44.652 & $-$1.04\,$\pm$\,0.15 & $-$2.86\,$\pm$\,0.16 & 269\,$\pm$\,124 & 37.8 & 2416\,$\pm$\,299 & 0.35 & 0.2 & 4380\,$\pm$\,810\\ [0.7ex]  
 NGC\,4815 & 194.499 & $-$64.963 & $-$5.77\,$\pm$\,0.11 & $-$0.95\,$\pm$\,0.10 & 371\,$\pm$\,128 & 15.3 & 3791\,$\pm$\,667 & 0.42 & 0.2 & 2230\,$\pm$\,60\\ [0.7ex] 
 Pismis\,18 & 204.230 & $-$62.095 & $-$5.65\,$\pm$\,0.12 & $-$2.28\,$\pm$\,0.10 & 575\,$\pm$\,199 & 12.0 & 2981\,$\pm$\,407 & 0.35 & 0.2 & 6320\,$\pm$\,25\\ [0.7ex] 
 Trumpler\,23 & 240.216 & $-$53.546 & $-$4.18\,$\pm$\,0.12 & $-$4.73\,$\pm$\,0.10 & 708\,$\pm$\,244 & 16.4 & 2820\,$\pm$\,370 & 0.42 & 0.2 & 4960\,$\pm$\,65\\ [0.7ex] 
 Trumpler\,20 & 189.886 & $-$60.654 & $-$7.09\,$\pm$\,0.10 & 0.18\,$\pm$\,0.09 &  1862\,$\pm$\,643 &  20.5 & 3955\,$\pm$\,618 & 0.18 & 0.15 & 3930\,$\pm$\,100\\ [1ex]
 \hline

\multicolumn{11}{l}{\footnotesize{\textsuperscript{a} Adopted from \citet{newcantatgaudin20}, $\Delta logt=0.20$ (NGC\,6067, NGC\,6259), $\Delta logt=0.15$ for the remaining clusters.}}
\end{tabular}
\label{table:values}
\end{table*}

We obtained average proper motion values that agree within the uncertainties with the values published in the literature. The same applies to the distances, all between 2.2 and 3.9\,kpc, yet our central values are between $\sim$100 and 500\,pc larger. Interestingly, the difference in distance that we get somehow scales with the distance itself.

Our dataset allows us to study the dispertion of proper motions and parallax for each cluster. The first quartile of our data is highly dominated by the brightest sources with the Gaia DR2 astrometry, whereas the fainter near-infrared sources, those in the third quartile, rely on our VVV astrometry. The proper motion dispersion of the clusters in our sample ranges from $\sim$0.1 to 0.2\,mas\,yr$^{-1}$. At the cluster distances, such a dispersion translates into a 1D velocity dispersion between $\sim$1.3 to 2.0\,km\,s$^{-1}$. Those are in agreement with the internal 1D velocity dispersions below the 2.0\,km\,s$^{-1}$ determined from spectroscopy for Pismis 18 \citep{hatzidimitriou19}, Trumpler 23 \citep{overbeek17} and Trumpler\,20 \citep{donati14}. The intrinsic parallax dispersion of cluster members must correspond to a physically plausible depth \citep{cantatgaudin20}. However, the parallax distribution of members of distant clusters is dominated by errors which, combined with the possible contamination of field stars, can artificially increase the observed proper motion dispersion. 

Using the tidal radius $r_t$ as a proxy to clusters concentration,  we find that the clusters of our sample are remarkably spatially concentrated. The youngest cluster ($\sim$125\,Myr) seems to be gravitationally-bound, while even Trumpler\,20, which is old and dynamically-evolved, is not very much physically extended (see $r_t$ values in Table\,\ref{table:values}). 

Due to the levity of interstellar extinction in the near-infrared, we can in this study map the cluster sequences better than the previous work based on optical data, and we report an increase of $\sim$45\% on average of cluster population. The improvement in the cluster’s quality near-infrared sequences is evident compared with the available near-infrared pass-bands of the 2MASS survey. Those, and our refinement of the clusters distances, allow us to updated qualitative isochrone fitting procedure.

\begin{figure*}
\includegraphics[scale=0.12]{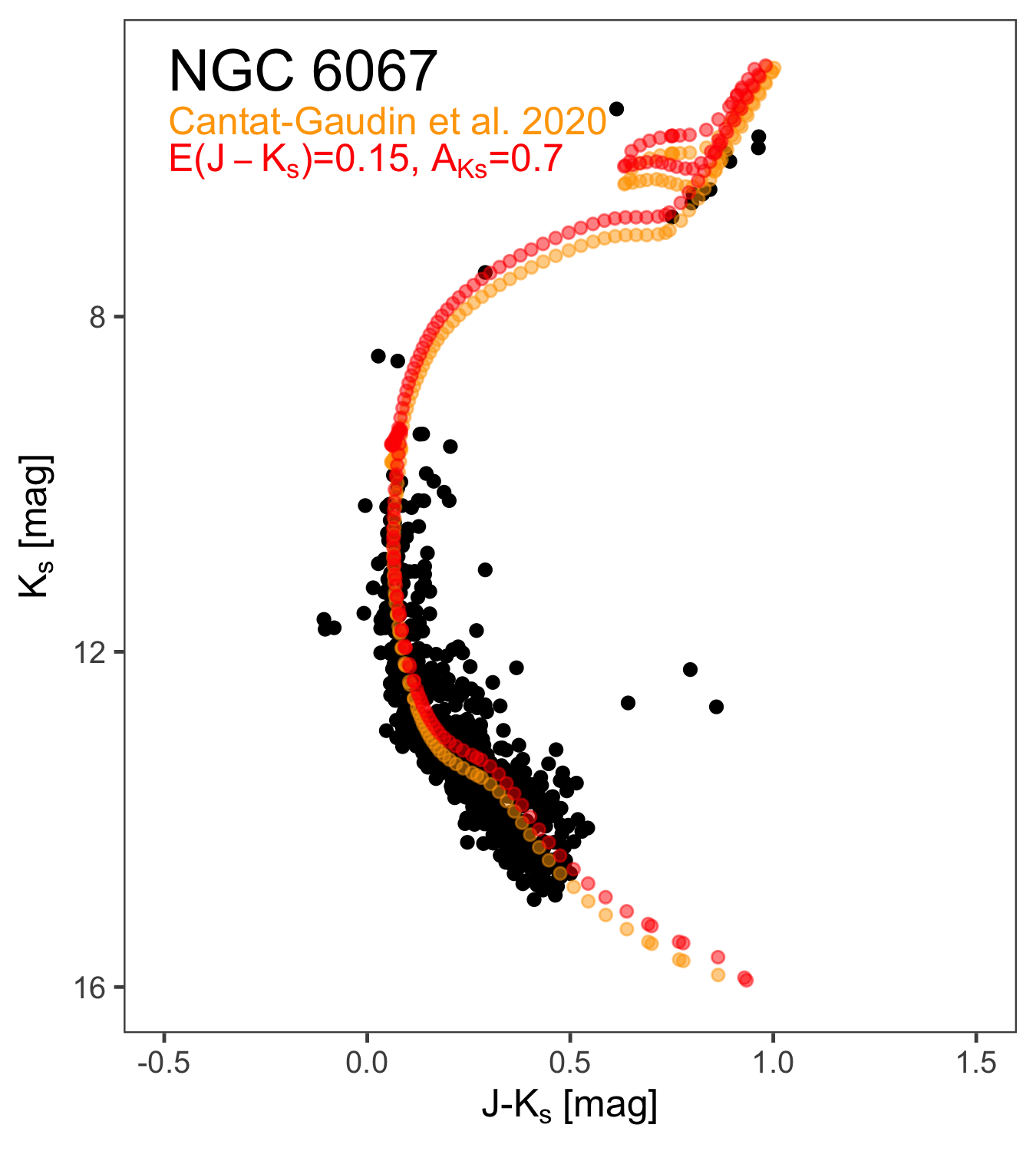}
\includegraphics[scale=0.12]{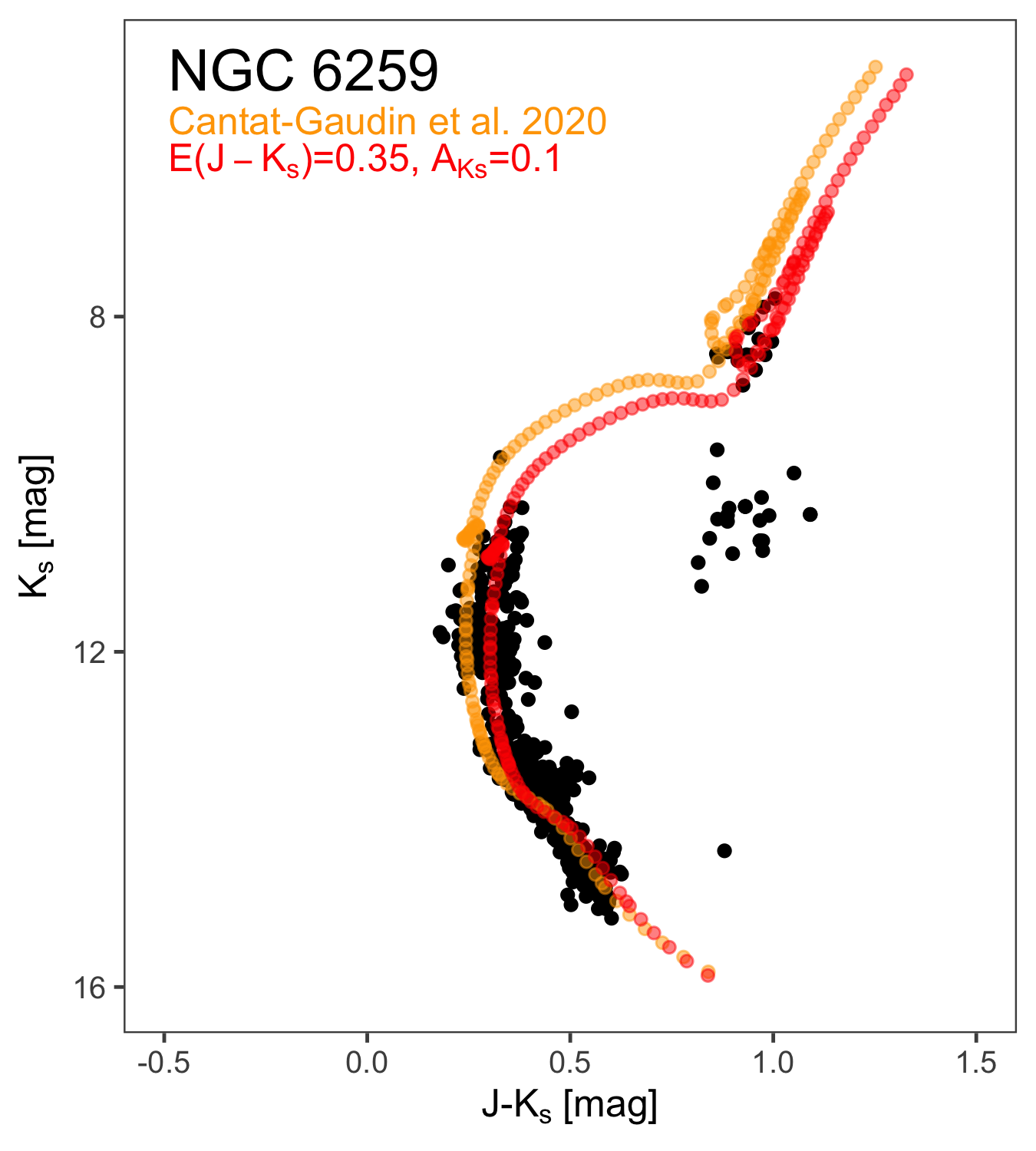}
\includegraphics[scale=0.12]{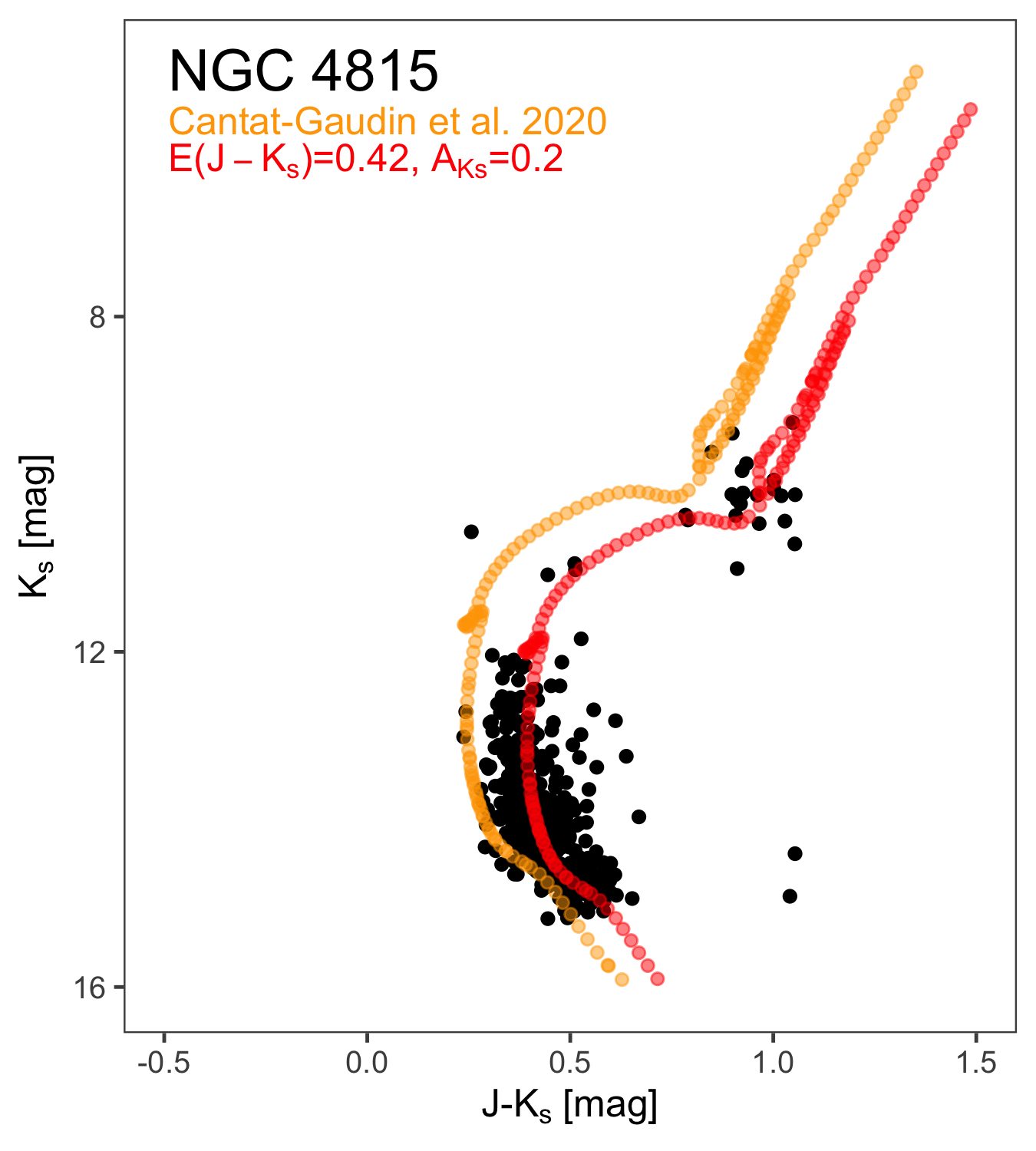}\\
\includegraphics[scale=0.12]{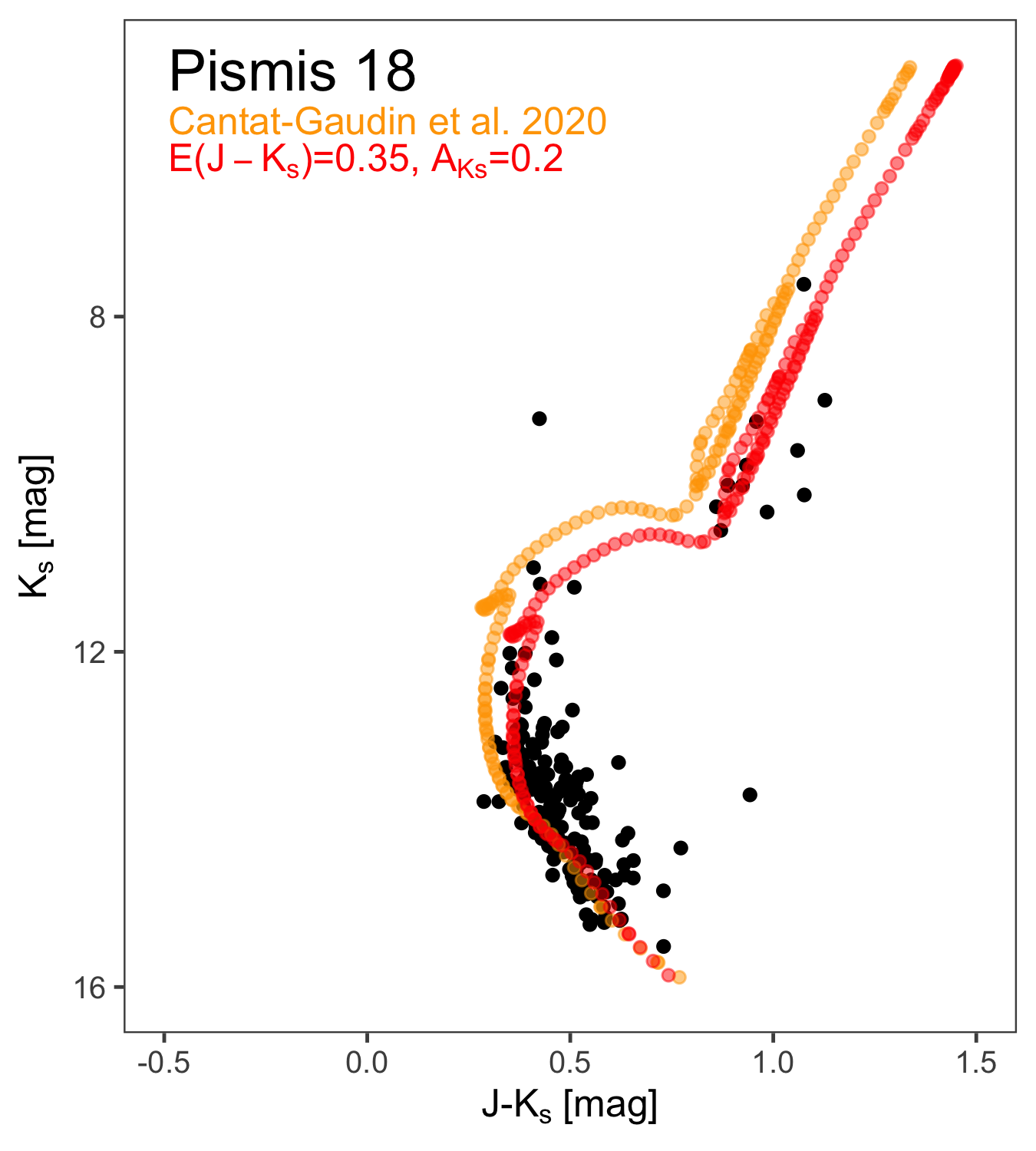}
\includegraphics[scale=0.12]{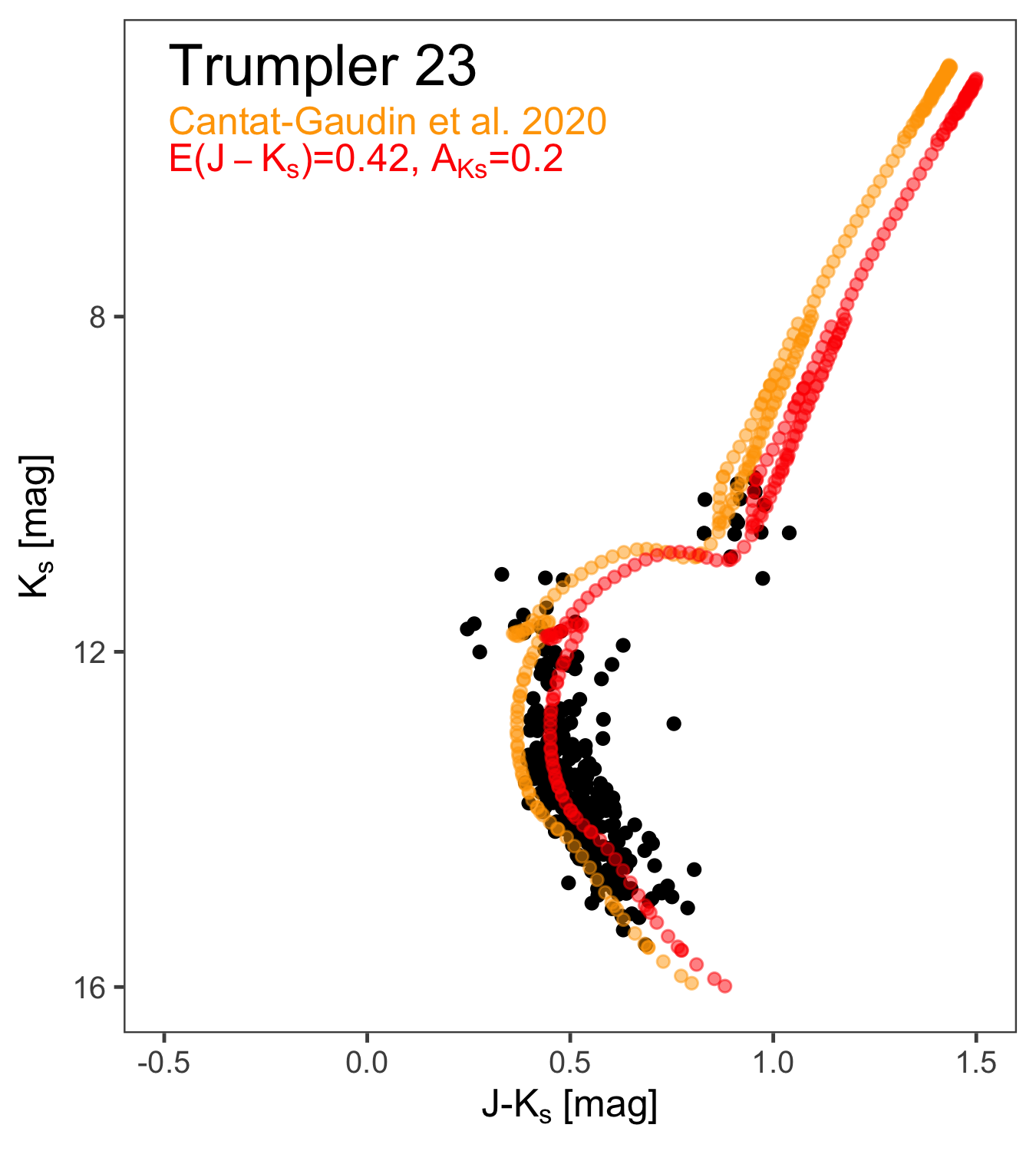}
\includegraphics[scale=0.12]{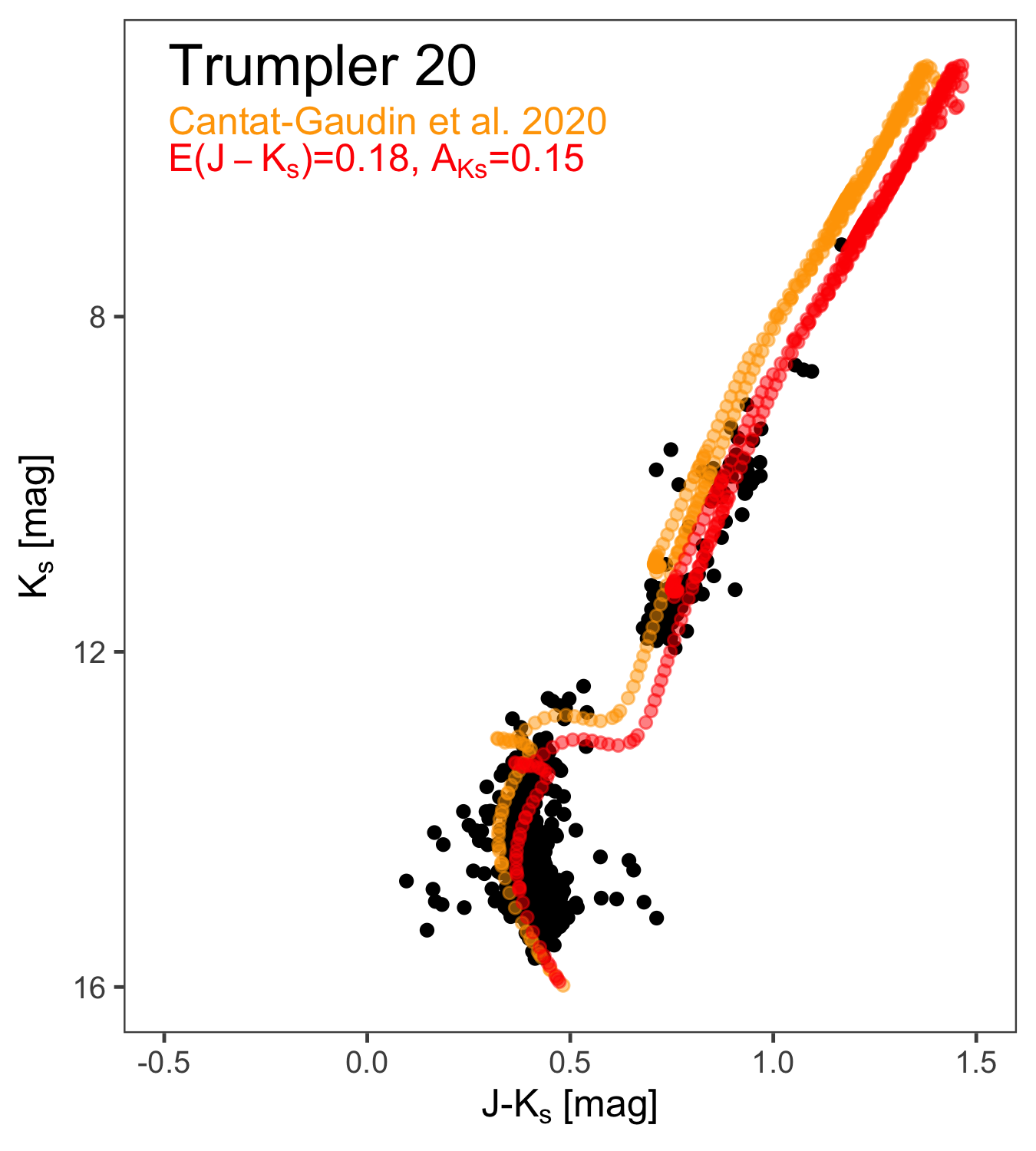}\\
\caption{Color-magnitude diagrams of the studied open clusters (absolute $K_s$, $J$-$K_s$). The \textsc{PARSEC-COLIBRI} isochrones are overplotted. The orange isochrones are located following the values of distance and extinction of \citet{newcantatgaudin20}. The red isochrone was shifted in each case to reproduce the high-probable near-infrared sequence and the derived values are reported along with each cluster label.}
\label{fig:iso}
\end{figure*}

In Figure\,\ref{fig:iso} we compare the absolute sequences in the VISTA photometric system to the models of \textsc{PARSEC-COLIBRI} \citep{marigo17}. For each cluster, we selected the model corresponding to the age derived in an homogeneous automated manner by \citet{newcantatgaudin20}, and applied the distance modulus and the reddening vector based on the distance and the median extinction $A_V$ from that same study. The reddening vector was transformed to $A_{K_s}$ using $A_{K_s}/A_V=0.12032$\footnote{\url{http://stev.oapd.inaf.it/cgi-bin/cmd}}), assuming the total-to-selective extinction ratios $R_J$ and $R_{K_s}$ of \citet{gonzalez18} for VISTA data. Since the resulting isochrones (plotted in orange in Figure\,\ref{fig:iso}) are not well aligned with our observed sequences, we also present the isochrones using our distance determination and adjusting the reddening. The adjustment was done using only the median of the 50\% of the highly probable members (p\,$>$\,90\%) that have the lowest reddening. This correction is presented with red isochrones in Figure\,\ref{fig:iso}. Our derived reddening and extinction values are presented in Table\,\ref{table:values}.

 Our improved values are an illustration that, as stated in \citet{platais12}, homogeneous data provide the best (internally consistent) parameters. No matter how good they are, if more than one dataset is used, the quality of the results is limited by systematic errors. One can note that even the fainter end of the sequences match the theoretical isochrones. Since we are still in the stellar regime at those magnitudes, we do not expect deviations due to the uncertainties in the models. That is an indication for a low rate of interlopers across the whole magnitude range covered in this study.

\subsection{Re-identification of known members}
Our membership determination relies on prior information on the central location, expected distance range, and an initial prior apparent size of the studied clusters. The radius we used around the clusters NGC\,6067, NGC\,6259 and, NGC\,4815 (defined as five times their $r50$ value from \citealt{cantatgaudin18}) includes other known clusters and cluster candidates. The area around NGC\,6067 includes FSR\,1716, and Harvard\,10, the area around NGC\,6259 overlaps with NGC\,6249, and the area around NGC\,4815 includes sources identified as members of Gulliver\,59. Our cluster sequence determination could successfully weed out all the contaminants and re-identify all the already known members of our  clusters. The following presents the comparison of our results with those in the literature. 

\begin{figure*}
\includegraphics[scale=0.09]{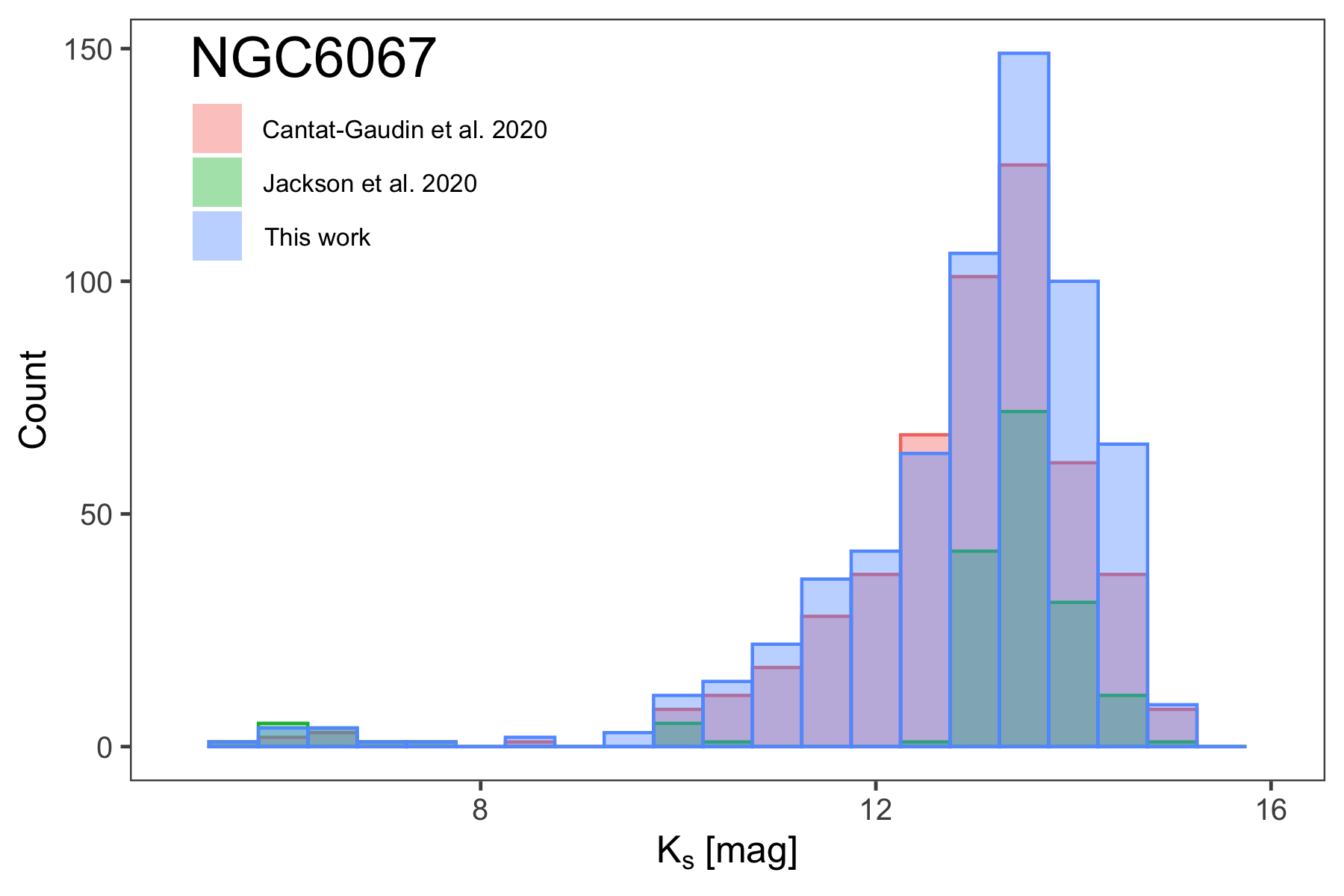}
\includegraphics[scale=0.09]{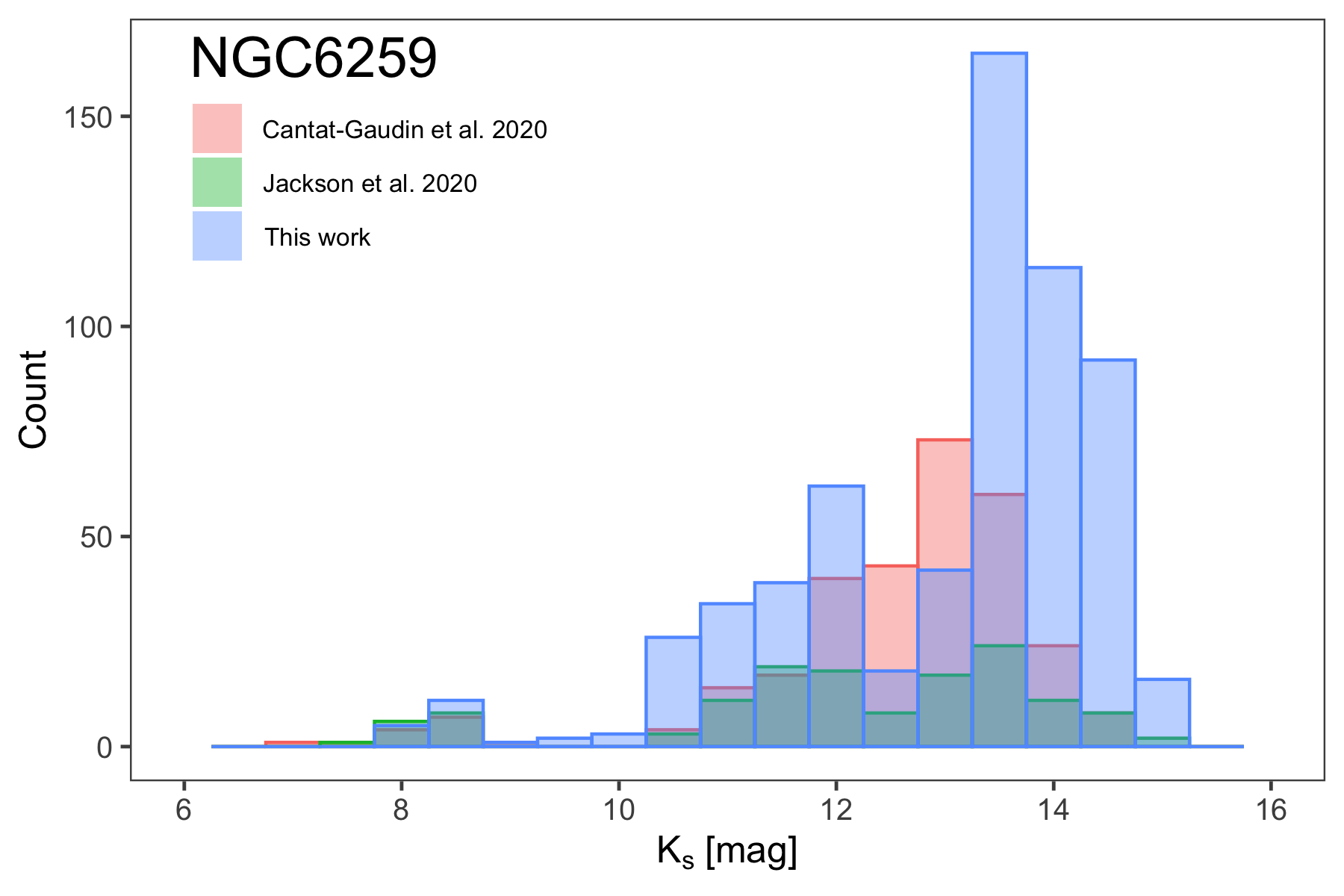}
\includegraphics[scale=0.09]{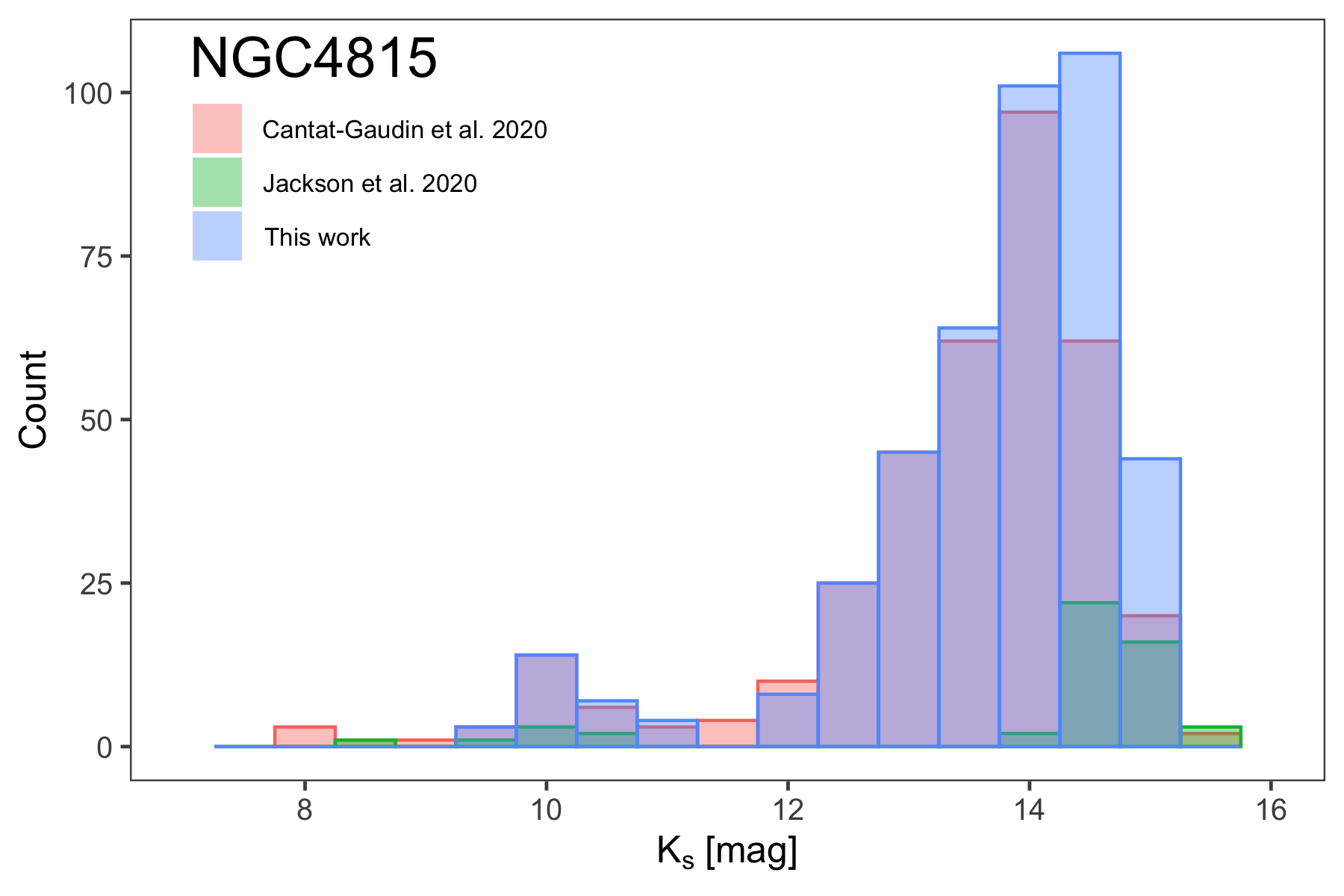}\\
\includegraphics[scale=0.09]{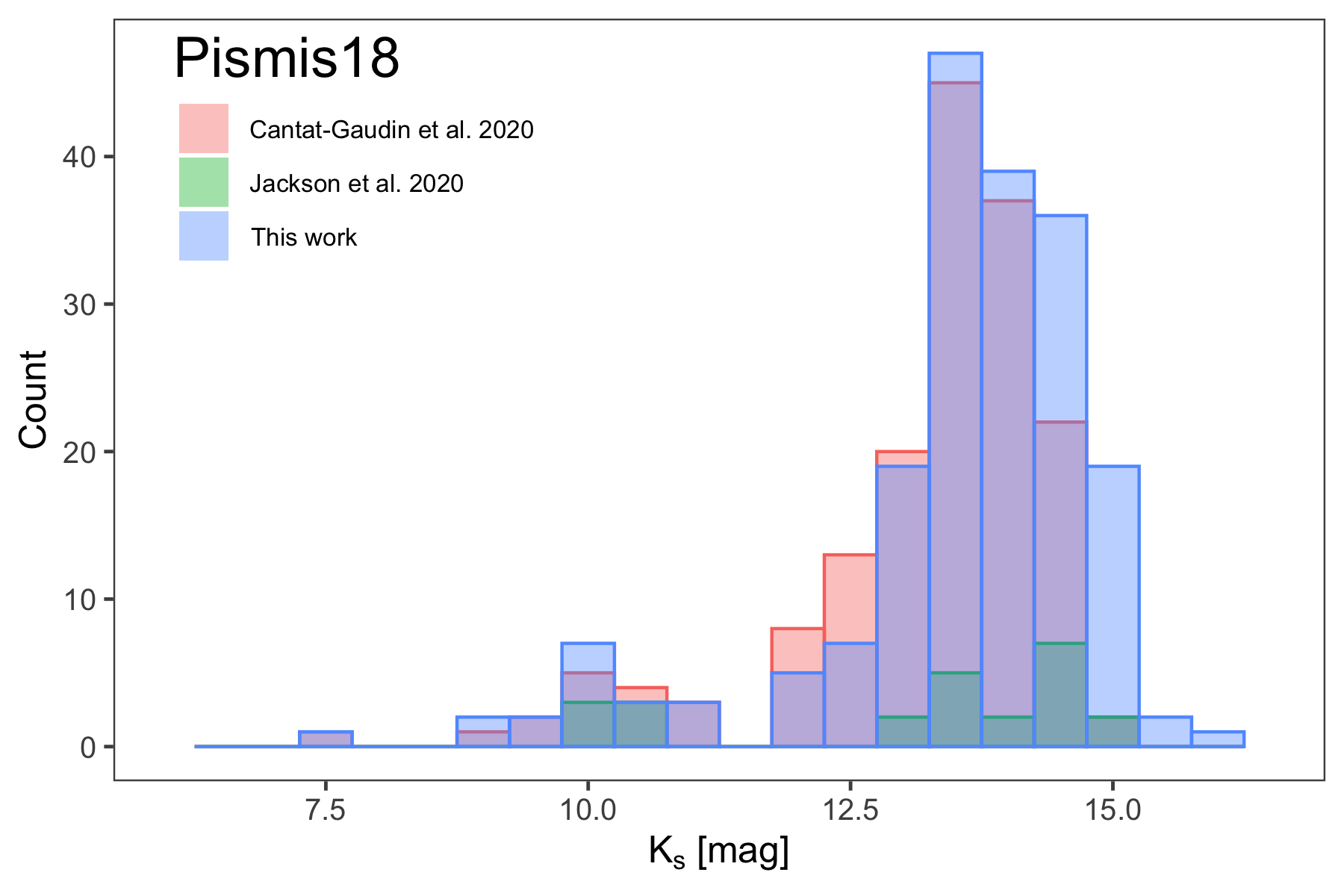}
\includegraphics[scale=0.09]{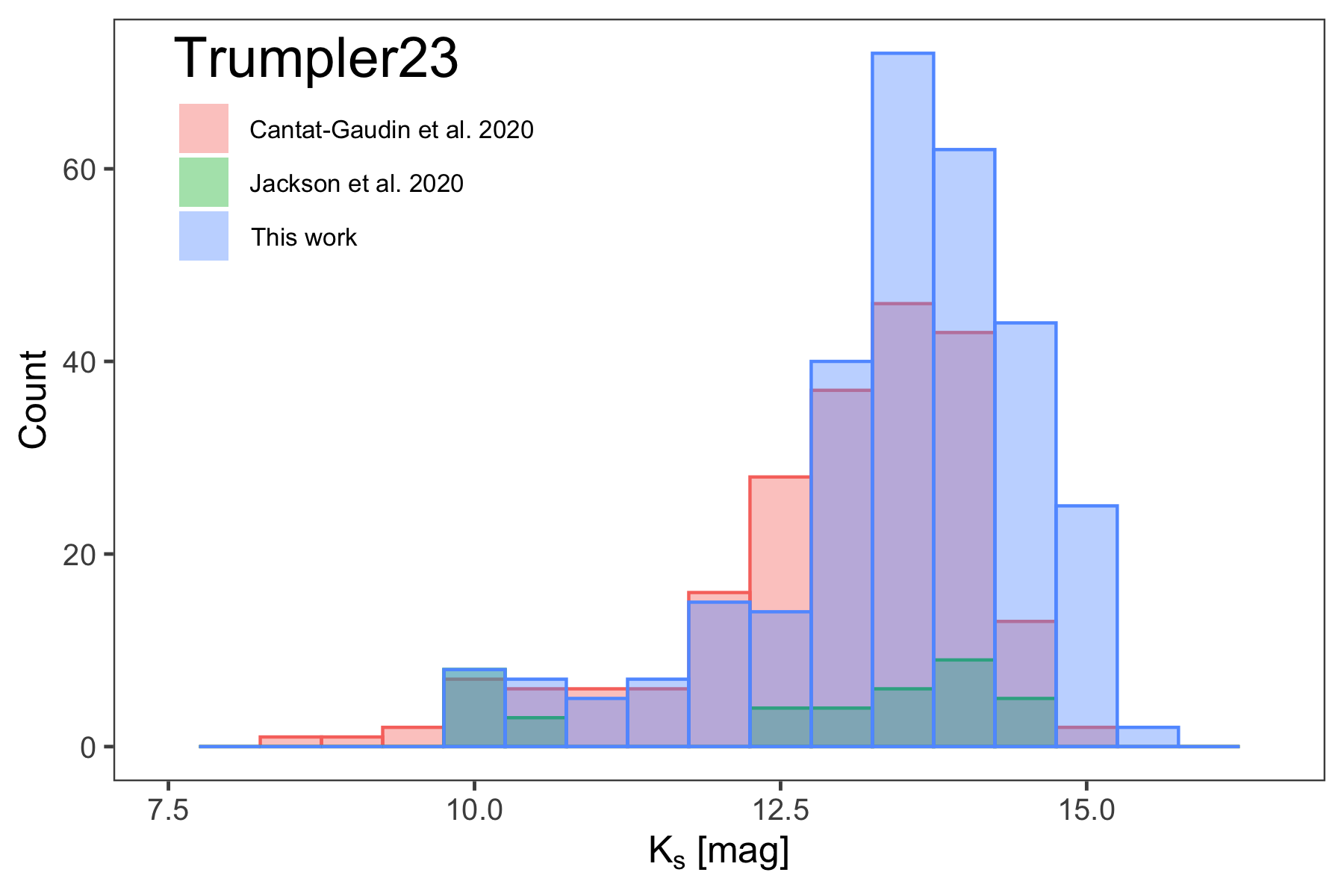}
\includegraphics[scale=0.09]{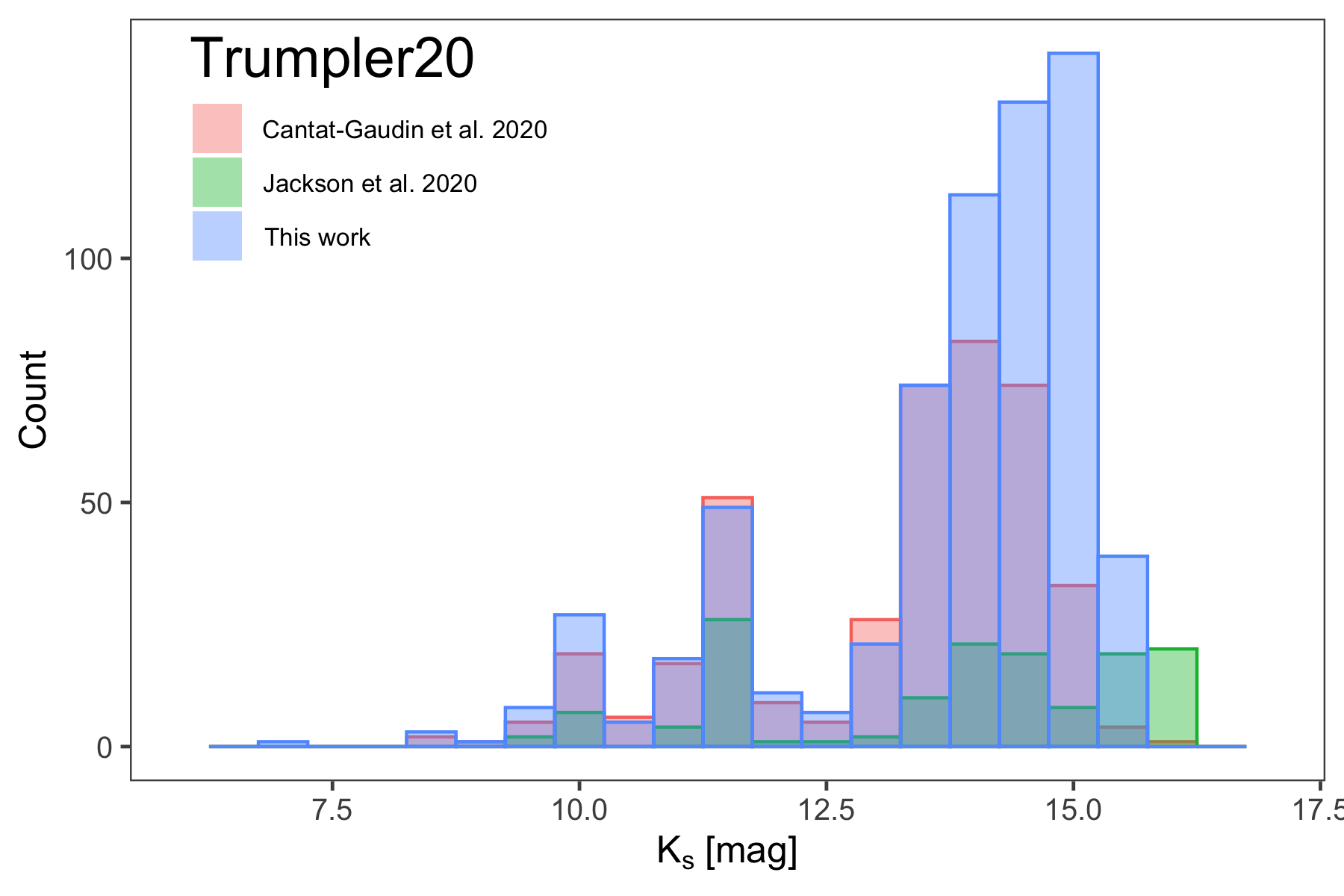}\\
\caption{$K_s$ magnitude distributions of the sources classified as members with probabilities $p\geq$90\% in the \citet{cantatgaudin20} and \citet{jackson20} studies, compared with our high probable members ($p\geq$90\%). Text on the plots shows their corresponding name. }
\label{fig:clusters_comp_lit}
\end{figure*}

For the broad comparison of our results with the literature, we have focused on the most recent works on the studied clusters, and specific notes on the clusters will be discussed in Section\,\ref{sec:notes}. We must first point out that our membership probabilities have a mean in the range 66--94\%, which is sightly higher than that of \citet[][with mean probabilites of 38--67\%]{cantatgaudin20}, or \citet[][with mean probabilites of 31--54\%]{jackson20}. With our approach, we are not recovering most of the low-probable candidates of those same studies. The Figure\,\ref{fig:clusters_comp_lit} presents the distribution of the $K_S$ magnitude of the highest probable members (p$\leq$90\%). From that figure, we can see that we recover a median of 77\% the high probable members of \citet{cantatgaudin20}, and 62\% of the high probable members of \citet{jackson20}. A great fraction of the new members are in the faint end of the near-infrared magnitudes. The non-recovered members are mainly sources that either exceed our search radius ($r_t$), were excluded by the GMM procedure, or did not pass our distance restrictions. Most of them have large proper motions and parallax uncertainties and are systematically skewed towards fainter magnitudes in the literature. It is worth mentioning that the distance ranges reported by \citet{cantatgaudin20} are shallower than those of \citet{jackson20} within their systematic uncertainties.

In NGC\,6259, we identify a group of high probability members at $K\sim$11\,mag that was missed by \citet{cantatgaudin20}, yet we failed to recover most of the members they found at $K\sim$12.5\,mag (see specific details in Section \ref{sec:notes}). The reason for this is still unclear, although this happens close to the boundary between VVV and 2MASS data, where our dataset is the most vulnerable to systematic errors. 

\subsection{Differences between datasets}
\begin{figure*}
\includegraphics[scale=0.14]{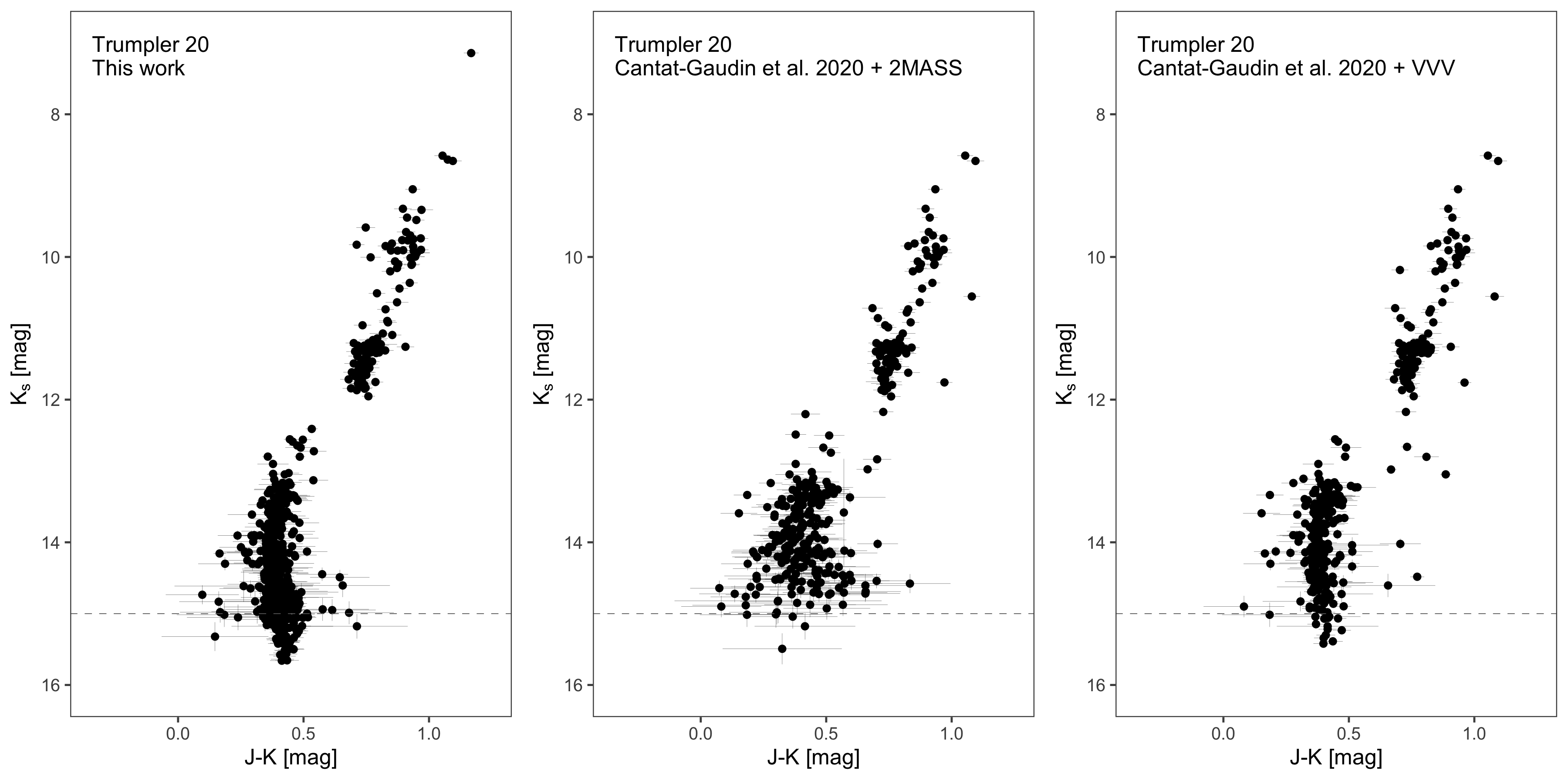}
\caption{Comparison of $K_s$ vs. $J-K_s$ color-magnitude diagrams for Trumpler\,20. \textit{Left:} Our VVV data following the procedure outlined in Section\,\ref{gmm_upmask}. \textit{Middle:} Members from \citet{cantatgaudin20} correlated with 2MASS data. \textit{Right}: Members from \citet{cantatgaudin20} directly correlated with our VVV photometry. The horizontal dashed line shows the depth of the cluster sequences at $K_s=15$\,mag. Only members with probabilities $p\geq$90\% are shown.}
\label{fig:t20_comp}
\end{figure*}

In Figures \ref{fig:t20_comp} and \ref{fig:all_comp}, we compare different datasets analyzed using the same methodology. For efficiency and to avoid lengthening this section too much, we only detail here the case of Trumpler\,20 -- the most distant/oldest cluster of our sample -- as an example, as it presents the same general trends than those seen for all the clusters in this study. The improvements are twofold; firstly, the deep VVV photometry provides a sharper clusters' sequence, avoids the contamination of red giants, and reaches deeper magnitudes. Secondly, even though the near-infrared sequence shows the same general trends as the optical one, there are differences introduced by our near-infrared native, data-driven approach. For example, in the case of Trumpler\,20, the evolved population ($K_s\lesssim12.0$\,mag) census increased by 25 newly identified stars, adding to the 99 ones isolated by \citet{cantatgaudin20} that we could confirm. At $K_s\sim$12.6\,mag we identify a group of sources that can be traced to an extended main sequence turn-off \citep{bastian18, dejuanovelar20} or to a metallicity gradient across the cluster main sequence turn-off. If confirmed, we lack of a specific mechanism to explain the presence of this population, although scenarios such as multiple populations and stellar rotation are suitable hypothesis. It is possible that this population has been missed in previous studies due to selection biases, and can only be brought into the light once a systematic exploration of proper motions and parallaxes is applied. As outlined in \citet{hoppe20}, stellar cluster membership is often based not only on stellar kinematics, but also on the assumption of an unique chemical abundance. In this sense, the samples devoid of individual objects with abnormal abundances by construction. In our case these populations are bright enough so that they are unlikely to be potential interlopers.

\subsection{Cluster total mass}
Each cluster's density profile was integrated to obtain the total number of stars contained in the cluster's main sequence. We determined the completeness magnitude for each cluster as the faintest magnitude at which the number of sources per interval of magnitude does not deviate from an increasing distribution. Assuming the central values of Kroupa mass function \citep{kroupa01} and the best-fitting PARSEC-COLIBRI isochrone, we extended the counting, adding all the masses of stars down to the stellar/substellar frontier (0.08\,M$\odot$). The evolved sources out from the main sequence were also linked to the best-fitting position on the corresponding isochrone. The mass values were weighted by the membership probability of each source.

The six studied clusters have total masses in the range 2230 $-$ 6320\,M$\odot$, and the values are consigned in Table\,\ref{table:values}. The adopted uncertainties consider the density profiles with and without weighting the mass values by each source's membership probability.  When we consider those uncertainties, the masses extreme values become 2170 $-$ 6340\,M$\odot$. In the case of NGC\,6259 the secondary group of sources at $(J-K_s)\sim1.0$\,mag, $K_s\sim$11\,mag (see Section \ref{sec:notes}) was isolated for its total mass determination. The total mass cluster values available in the literature show a significant dispersion. For NGC\,6067, the extreme values goes from 893 to $\sim$5700\,M$\odot$ \citep{piskunov08, alonsosantiago17}, while we report an intermediate value. In the case of Trumpler\,20, we report a total mass deviating at a 3$\sigma$ level from lowest mass of the values reported in the literature: 6700\,$\pm$\,800\,M$\odot$ \citep{donati14} and 15000\,M$\odot$ \citep{gunes17}. In the case of NGC\,4815, we present a total mass value that duplicates more consolidated values based on optical photometry data \citep{chen98, prisinzano01}. Finally, \citep{bonatto07} reported a total mass value of 3100\,$\pm$\,1300\,M$\odot$ for Trumpler\,23 which is at 1.4$\sigma$ from the total mass value presented here. To the best of our knowledge, there are no previous total mass estimates for the clusters NGC\,6259 and Pismis\,18.

\section{Notes on specific clusters}
\label{sec:notes}
\subsection{NGC\,6067}
This cluster has a diverse population composed of B-type dwarfs and giants, AF-type supergiants, and two Cepheids. Our catalog does not report the G1\,Iab Cepheid QZ Nor as member, since even though Gaia DR2 provides radial velocity and parallax that are consistent with its membership to the cluster, there is no proper motion information available. Also, the metallicity discrepancy reported for that star by \citep{alonsosantiago17} remains unsolved. 

Regardless of its proximity, the planetary nebula Pn\,HeFa\,1 is not associated to the cluster \citep{monibidin14}, although the high-mass planetary nebulae BMP\,J1613-5406 is \citep{fragkou19}. Using this object \citet{fragkou19} sets the cluster's main sequence turn-off point at the B6V-types, agreeing with \citealt{alonsosantiago17} and \citealt{mermilliod81}). From our analysis, we located it precisely at $\sim$4.6\,M$_\odot$.

\subsection{NGC\,6259}
There has been numerous attempts to characterize this cluster at optical wavelengths \citep{hawarden74,twarog89,mermilliod01,claria08}. It was also included in the works of \citet{magrini18no,magrini18} who were assessing its abundance pattern of s-process elements and its N/O  abundance ratio. Assuming an age of 0.2\,Gyr, \citet{lagarde19} derived a turn-off mass of 3.7\,M$_\odot$. From our analysis, we relocate the turn-off mass at $\sim$3.4\,M$_\odot$. The cluster population has been studied looking for binary systems with negative results \citet{mermilliod08}, while \citep{ogle06} claims eclipsing variables are associated to the cluster based on their color and magnitude. From our analysis, the source 2MASSJ17004708$-$4437175 is the only known variable star associated with NGC\,6259.

One can clearly see a group of sources at  $(J-K_s)\sim1.0$\,mag, $K_s\sim$11\,mag, that share common properties with the most probable cluster members. This group appears isolated in the CMD and is too faint to fit the isochrone. Given their proximity to NGC\,6259 these sources can not be members of NGC\,6249, although their spatial distribution in right ascension is asymmetric. Their radial velocities obtained from the literature are in complete agreement with the cluster's mean radial velocity. To assess the membership status of this group of sources we have repeated the same membership determination algorithms discussed in this work to the sources lying in an ring around the cluster and beyond $r_t$ (up to the equivalent in area). We did not recover a single source in the color-magnitude region where this group of sources linked to NGC\,6259 lies. We also attempted various isochrone fittings exploring the possibility of a difference in age but none of the values explored can link the faint group of sources to the cluster main sequence. We can conclude that it is unlikely that these sources at $(J-K_s), K_s\sim$1.0, 11.0 are interlopers from the foreground/background  or due to systematic errors on the methodology.

\subsection{NGC\,4815}
This open cluster has been extensively studied in terms of its C, N, O, Na, Al abundances and abundance ratios of elements belonging to different nucleosynthetic channels \citep{taut15,friel14,magrini14} mostly based on Gaia-ESO data. Its spatial distribution was explored in the optical by \citet{chen98} and based on their members selection the authors report a higher concentration of the cluster's brighter stars. With our access to fainter magnitudes, the effect seems less evident, especially given that our exploration reaches the cluster tidal radius. The luminosity function and stellar mass function for NGC\,4815 have been studied by \citet{prisinzano01,sagar01}, finding it consistent with a Salpeter slope.

\subsection{Pismis\,18}
The cluster has been recently revisited using the optical spectroscopic Gaia-ESO survey by \citet{hatzidimitriou19}. Their observational sample is based on the membership analysis of \citet{cantatgaudin18}. They illustrated the importance of using proper motions when assigning cluster membership. Indeed, there are several objects with radial velocities consistent with their membership to the cluster, but which are scattered over large distances from the cluster's center and have inconsistent proper motion values. The authors found a mean parallax different from literature, but the same survey team solved that issue in \citet{jackson20}.

\subsection{Trumpler\,23}
The study by \citet{bonatto07}, based on 2MASS photometry only, shows the analysis of this cluster to be challenging due to complications with crowding, differential reddening, and possible tidal effects. Later on, \citet{overbeek17} used Gaia-ESO data to isolate 70 sources members, based on their radial velocity. Their cluster parametrization agrees within the uncertainties with the values presented in our work, although the authors locate the cluster at a closer distance. \citet{overbeek17} found an apparent spread in the lower main sequence, which they explain with a broadening from both differential reddening and binaries, as well as the possible presence of a blue and a red main sequences. Our study does not confirm a double sequence, although there is an evident spread in the sequence (see Figure\,\ref{fig:iso}). From our membership assessment, there are two faint-red sources identified as probable members in Trumpler\,23 that can be potential interlopers given their high parallax errors. 

\subsection{Trumpler\,20}
This an old open cluster located in the inner disc (beyond the great Carina spiral arm) that has been targeted by the Gaia-ESO Survey \citep{donati14}, which sample was only limited to the cluster's core. As shown in Figure\,\ref{fig:clusters_info}, Trumpler\,20's sequence is broad, especially at faint magnitudes, and seems effected by differential reddening, which is consistent with its position in the Galactic disc and its reddening \citep{donati14, platais12}. The source of the differential reddening could mainly be a patchy dust structures in the field of view (see Figure\,\ref{fig:clusters}), along the line of sight. Using isochrone fitting of the turn-off, \citet{carraro14} derived cluster parameters that are in agreement with the most recent literature data and our derivations.

\section{Conclusions}
This study unveils the near-infrared sequence of six open clusters located up to $\sim$4\,kpc. This work employs a data-driven approach tailored to disentangle the cluster population from the field stars. The cluster members' search is centered in two main steps: a density-aware exploration of the data in the positional and astrometric space and a cluster membership assignment based on positional and photometric similarity. 

Our methodology has allowed us to disentangle the cluster populations from their dense backgrounds. Our study increased by $\sim$45\% on average the number of cluster members, building up our knowledge on the near-infrared cluster sequences. The calculated physical parameters of the clusters are in general agreement with the literature values, although we do update their extinction values and total masses by unveiling a large portion of the low-to-intermediate mass population with a native near-infrared data-driven approach.

There are about 150 clusters that \citet{newcantatgaudin20} could not analyse, either because they were too red or had not enough stars. Those clusters are now accessible with our framework. We use the unprecedented quality of astrometry provided by the Gaia mission as a starting point. Yet, we also include VISTA near-infrared photometry and astrometry to unveil the near-infrared counterpart of each studied cluster. Our approach's strength does not only reside in good quality parallaxes from which distances can be obtained and in precise and deep near-infrared photometry, but also in our ability to obtain clean near-infrared color-magnitude diagrams from which astrophysical parameters can be inferred.

\section*{Acknowledgements}
We are grateful to the referee, Giovanni Carraro, for helpful comments which significantly helped improve the paper. This work was supported by MINEDUC-UA project code ANT1855, CONICYT$-$ANID FONDECYT Regular 1201490, CONICYT$-$ANID FONDECYT Iniciaci\'on 11201161, 1171025, and CONICYT PAI ``Concurso Nacional Inserci\'on de Capital Humano Avanzado en la Academia 2017'' project PAI79170089. This paper made use of
the Whole Sky Database (wsdb) created by Sergey Koposov and maintained at the Institute of Astronomy, Cambridge by Sergey Koposov, Vasily Belokurov and Wyn Evans with financial support from the Science \&
Technology Facilities Council (STFC) and the European Research Council (ERC). This work was supported by the international Gemini Observatory, a program of NSF's NOIRLab, which is managed by the Association of Universities for Research in Astronomy (AURA) under a cooperative agreement with the National Science Foundation, on behalf of the Gemini partnership of Argentina, Brazil, Canada, Chile, the Republic of Korea, and the United States of America. This work has made use of data from the European Space Agency (ESA) mission
{\it Gaia} (\url{https://www.cosmos.esa.int/gaia}), processed by the {\it Gaia} Data Processing and Analysis Consortium (DPAC, \url{https://www.cosmos.esa.int/web/gaia/dpac/consortium}). Funding for the DPAC has been provided by national institutions, in particular the institutions participating in the {\it Gaia} Multilateral Agreement.

\section{Data availability}
\label{catalogs}
The data underlying this article are available in the Open Software Foundation, at \url{https://dx.doi.org/10.17605/OSF.IO/359QD}.

\bibliographystyle{mnras}
\bibliography{references} 

\appendix
\renewcommand\thefigure{\thesection.\arabic{figure}}
\section{Detailed color-magnitude diagrams.}
\setcounter{figure}{0}    

\begin{figure*}
\includegraphics[scale=0.12]{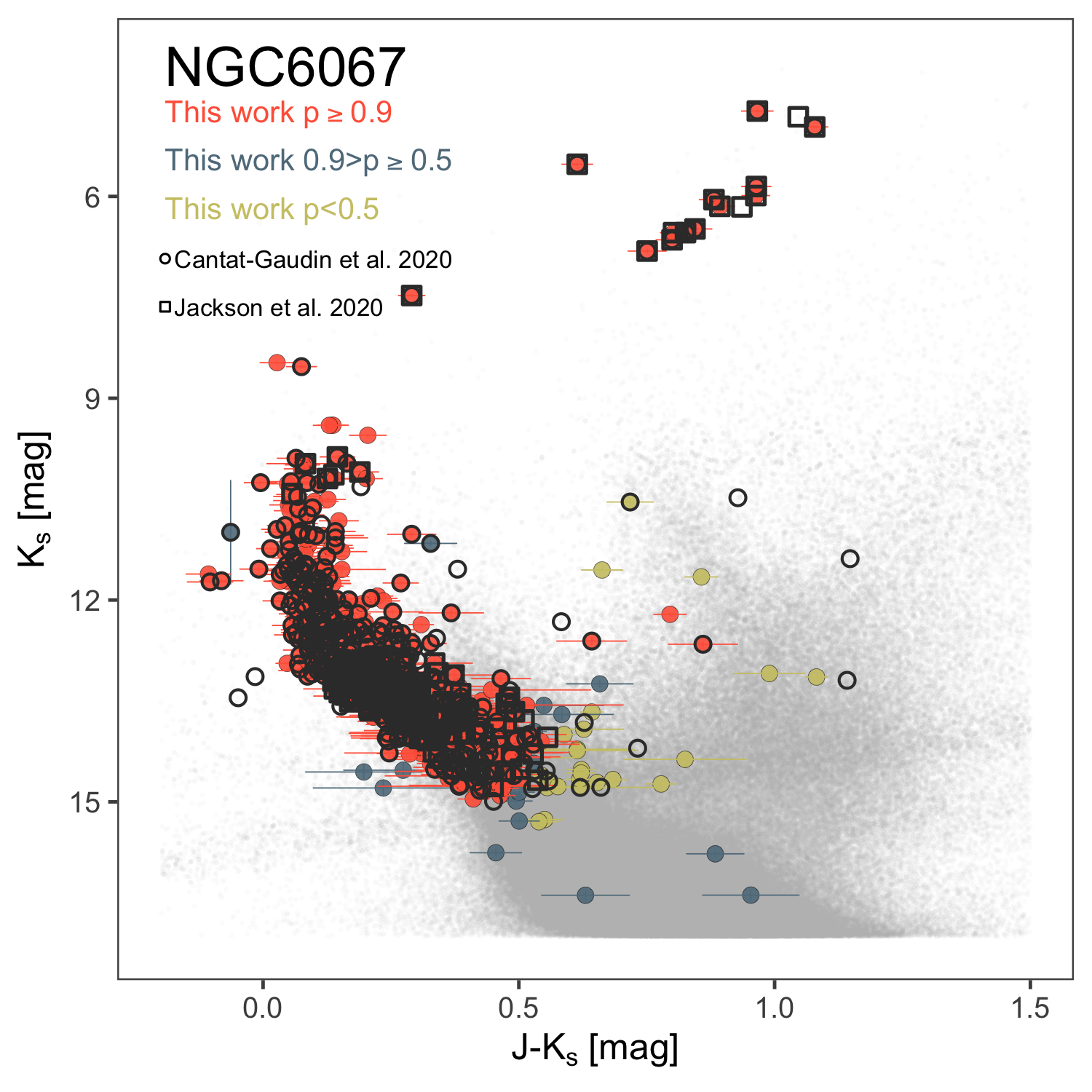}
\includegraphics[scale=0.12]{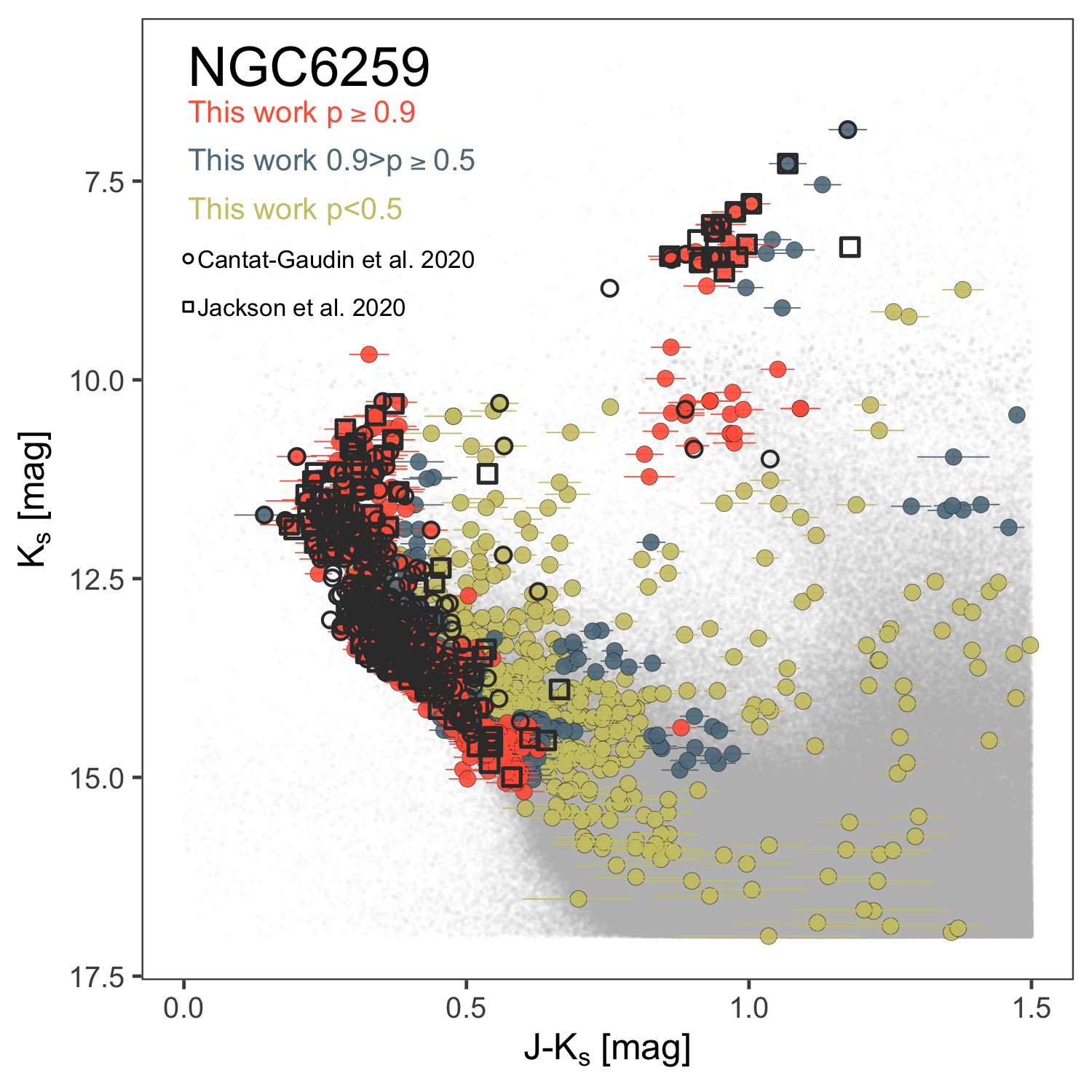}\\
\includegraphics[scale=0.12]{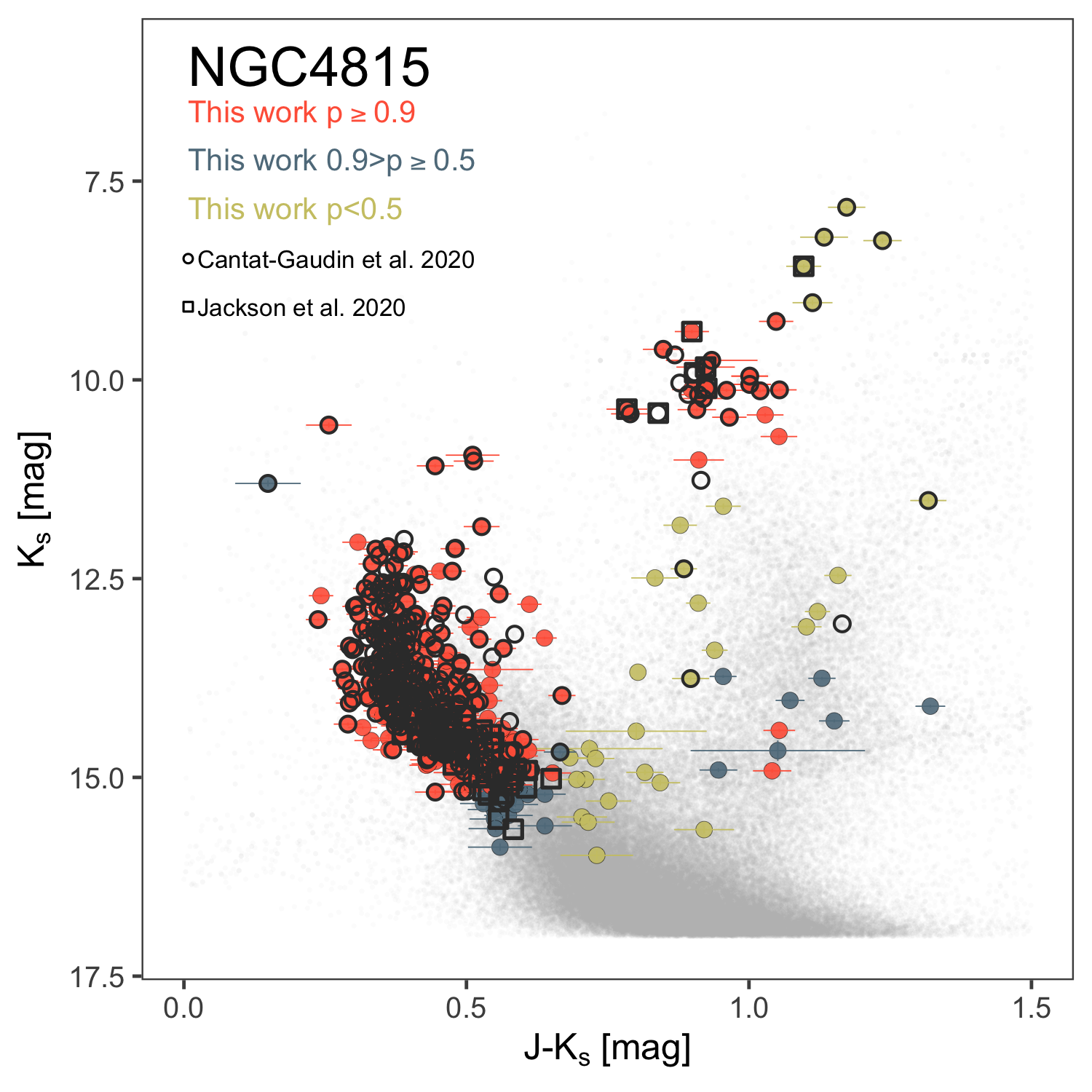}
\includegraphics[scale=0.12]{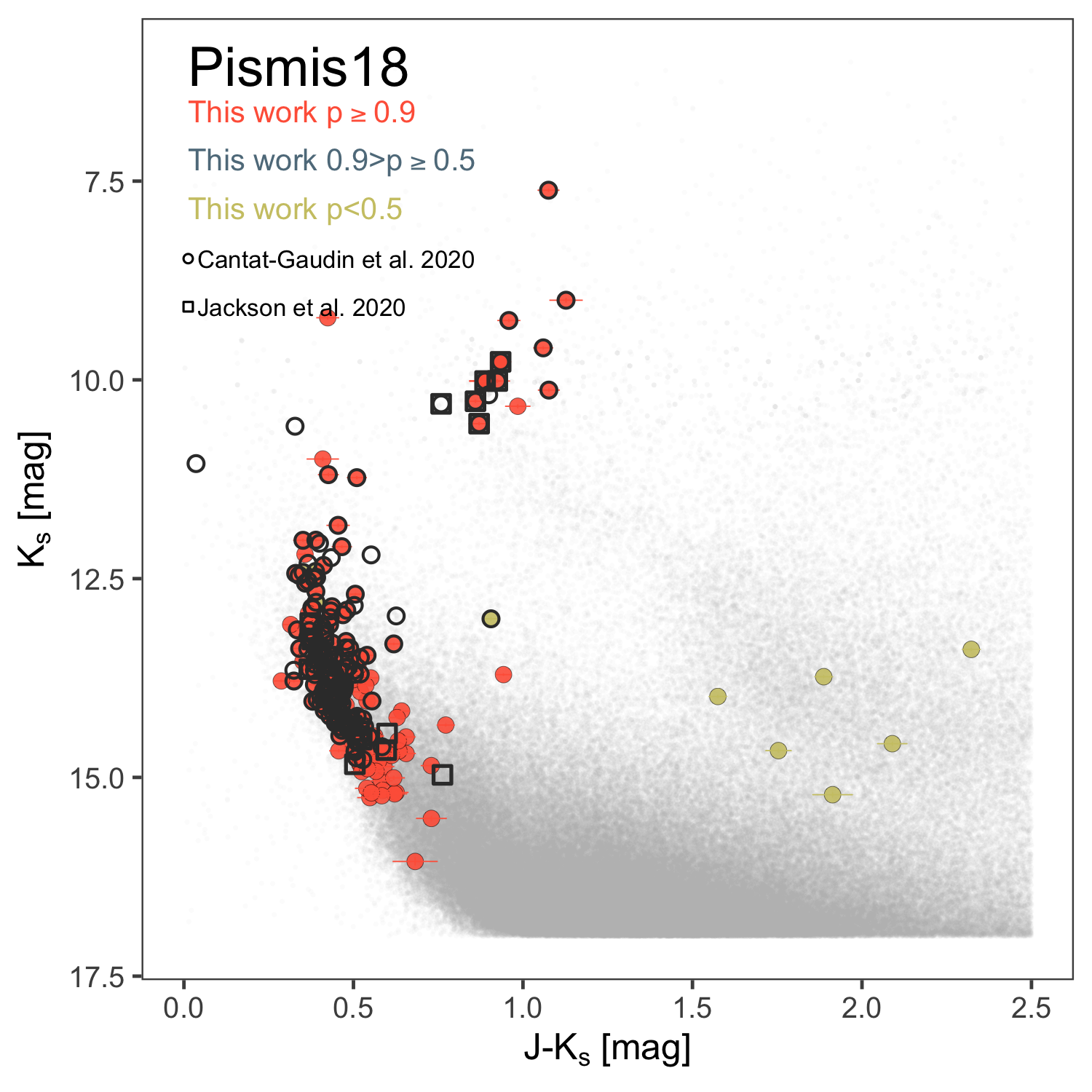}\\
\includegraphics[scale=0.12]{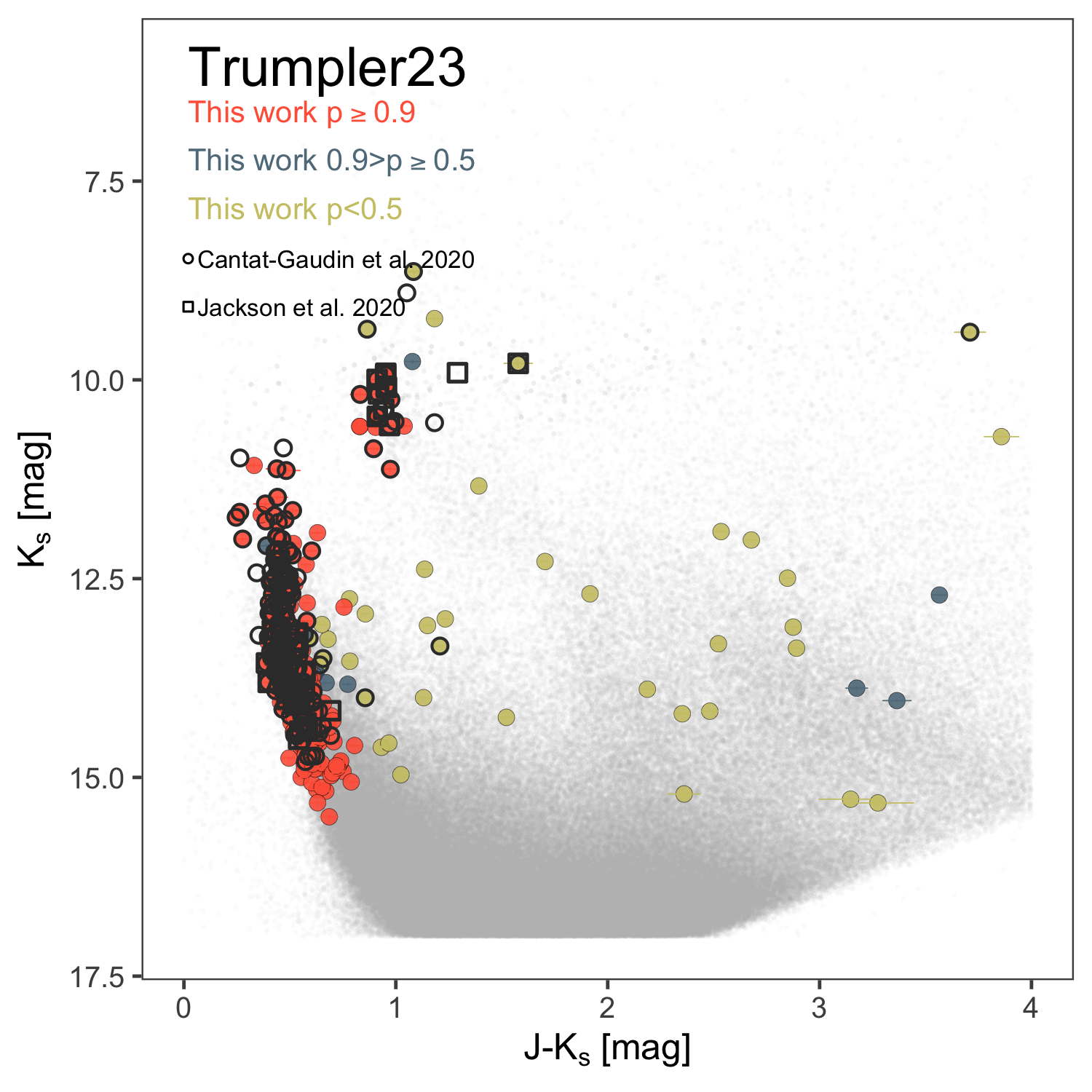}
\includegraphics[scale=0.12]{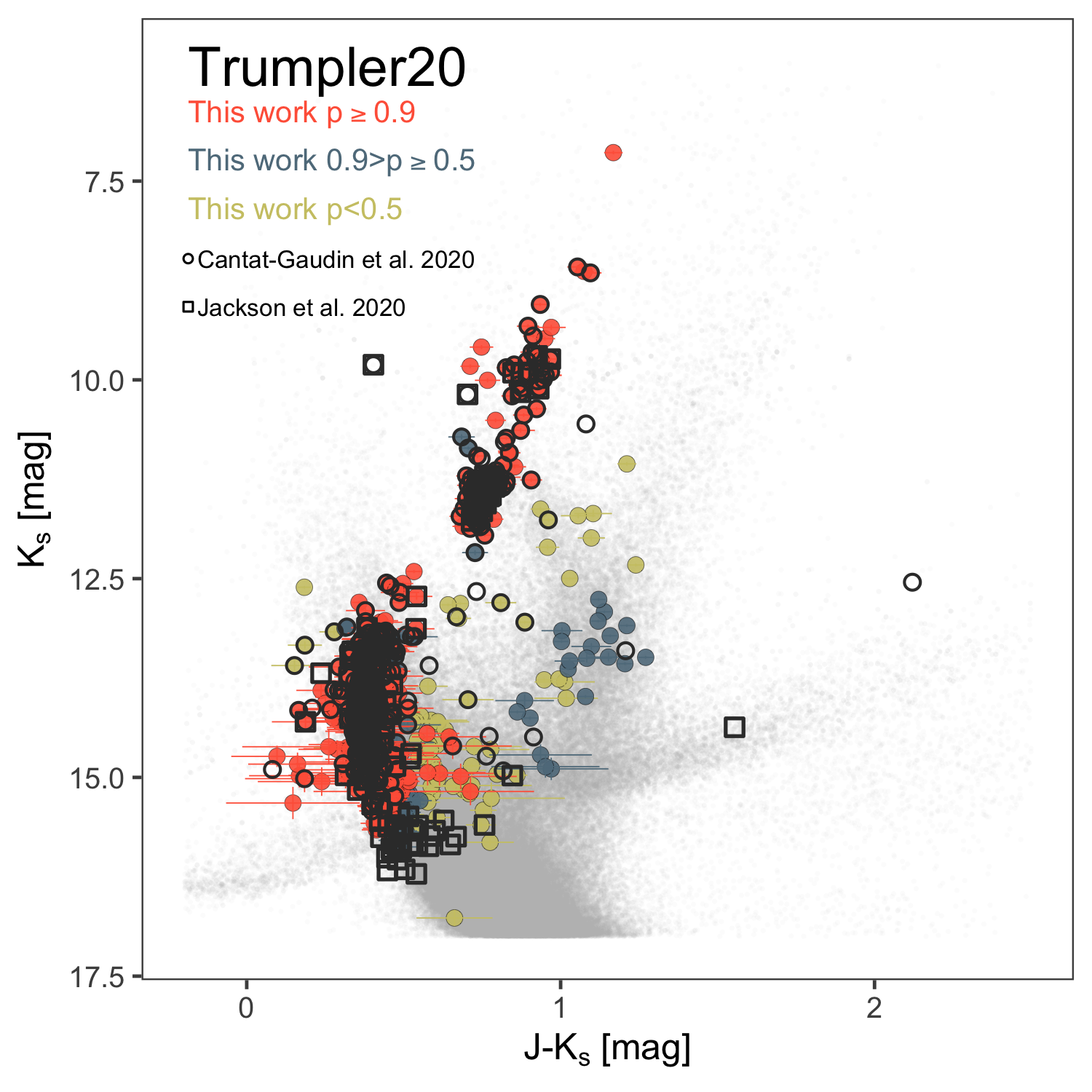}\\
\caption{Detailed $K$ vs. $J$-$K$ color-magnitude diagrams of the studied open stellar clusters. Cluster members with membership probabilities $p\geq$90\% are represented as red dots. Sources with membership probabilities between $0.9>$p$\geq0.5$ and p$<0.5$ are represented with dark blue and light green dots. Sources identified with high membership probability (p$\geq0.9$) from \citet{cantatgaudin20} and \citet{jackson20} are represented as open circles and open boxes, respectively. Field contaminants are presented as light gray dots. The two faint-red sources identified as probable members in Trumpler\,23 are not shown in previous figures. Text on the plots shows their corresponding name.}
\label{fig:clusters_info}
\end{figure*}

\begin{figure*}
\includegraphics[scale=0.07]{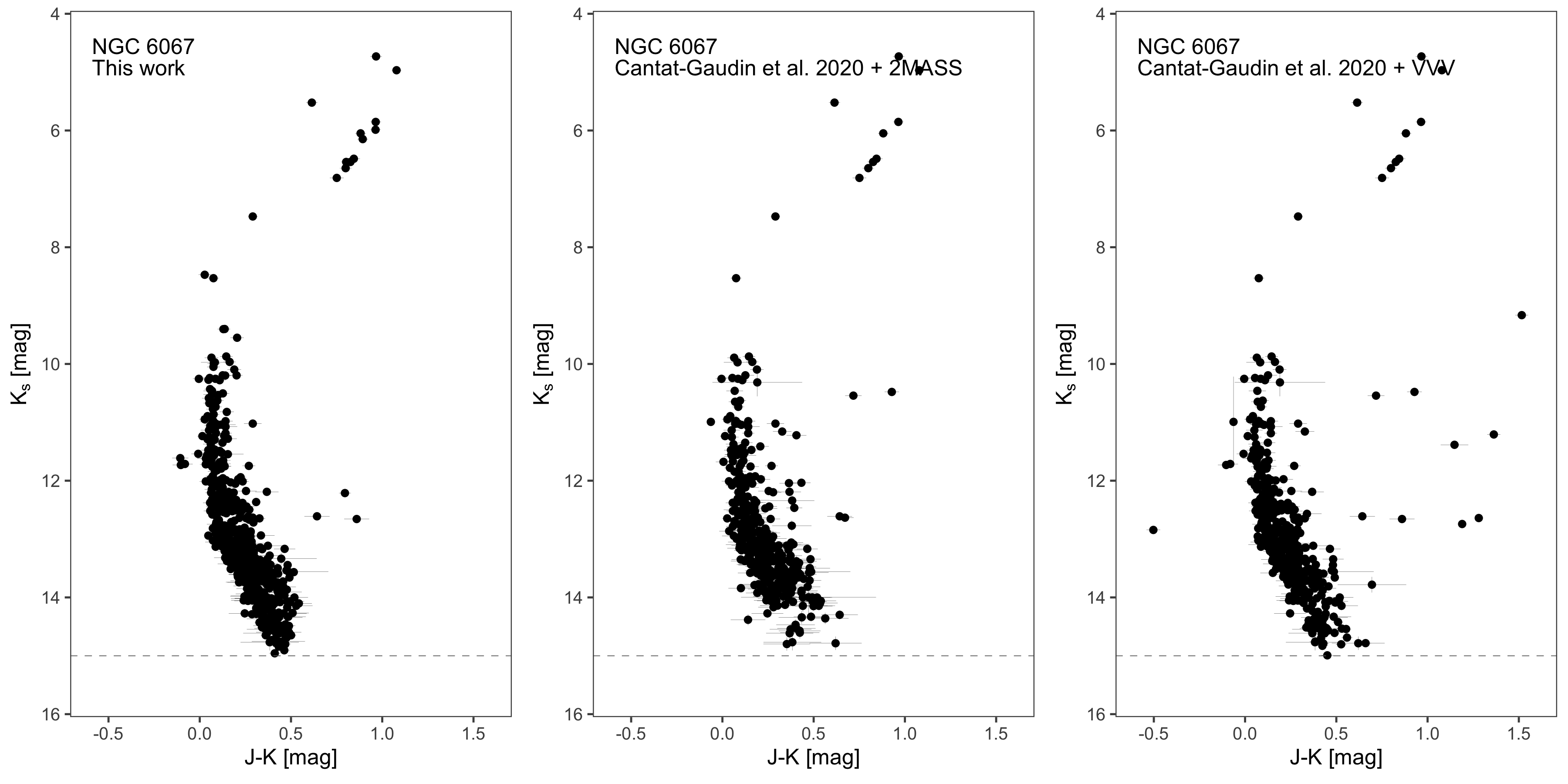}
\includegraphics[scale=0.07]{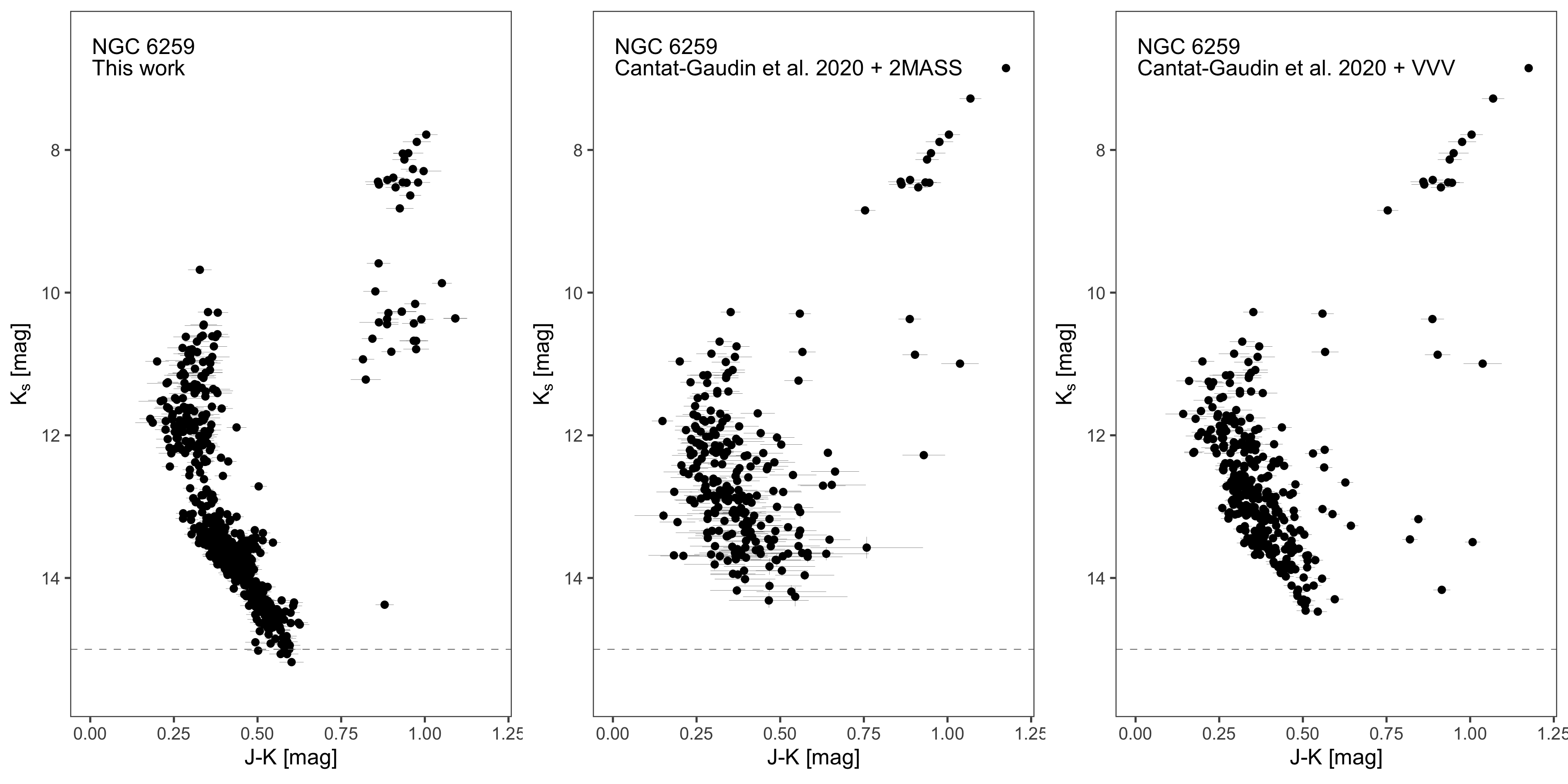}
\includegraphics[scale=0.07]{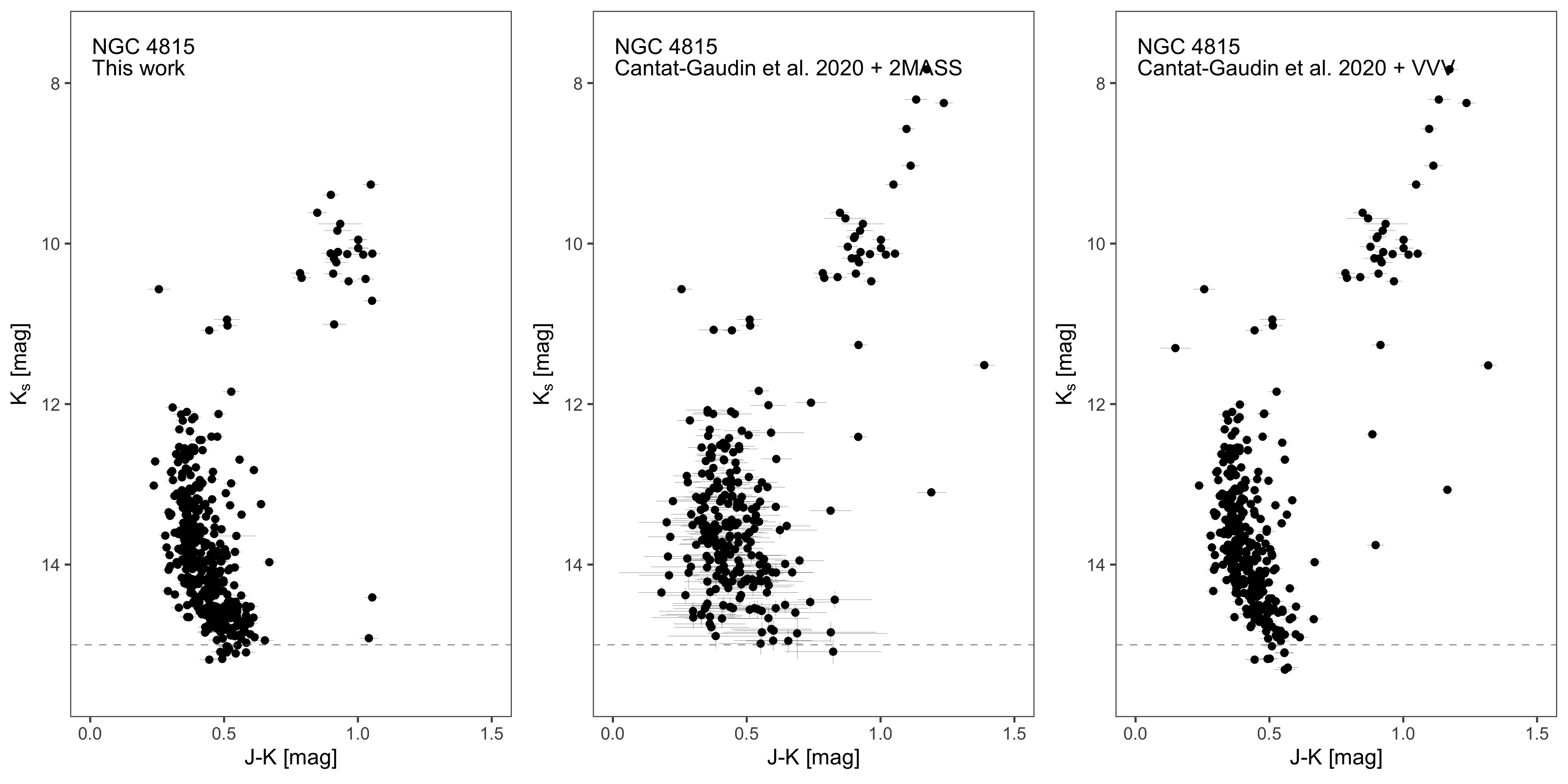}
\includegraphics[scale=0.07]{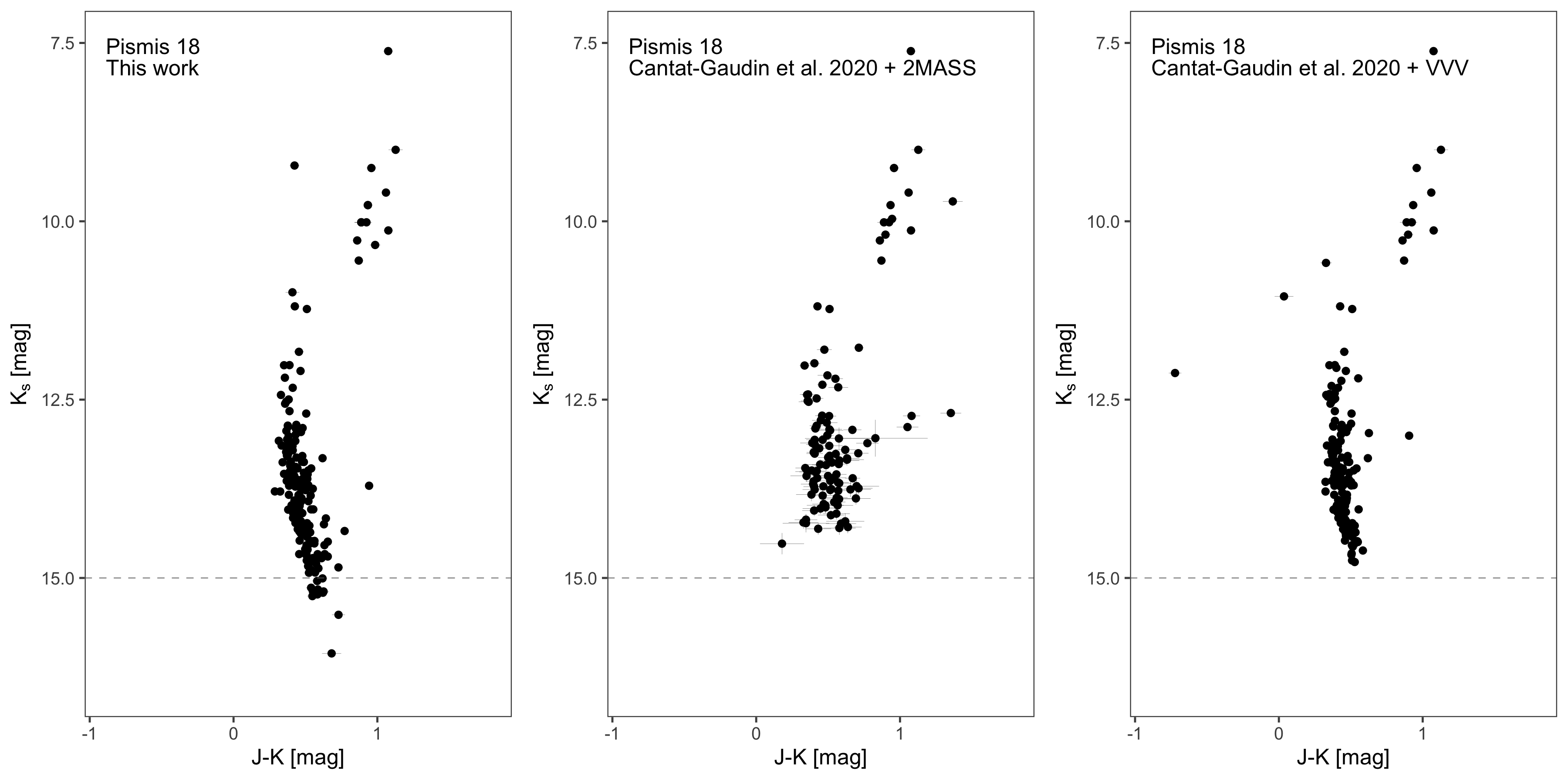}
\includegraphics[scale=0.07]{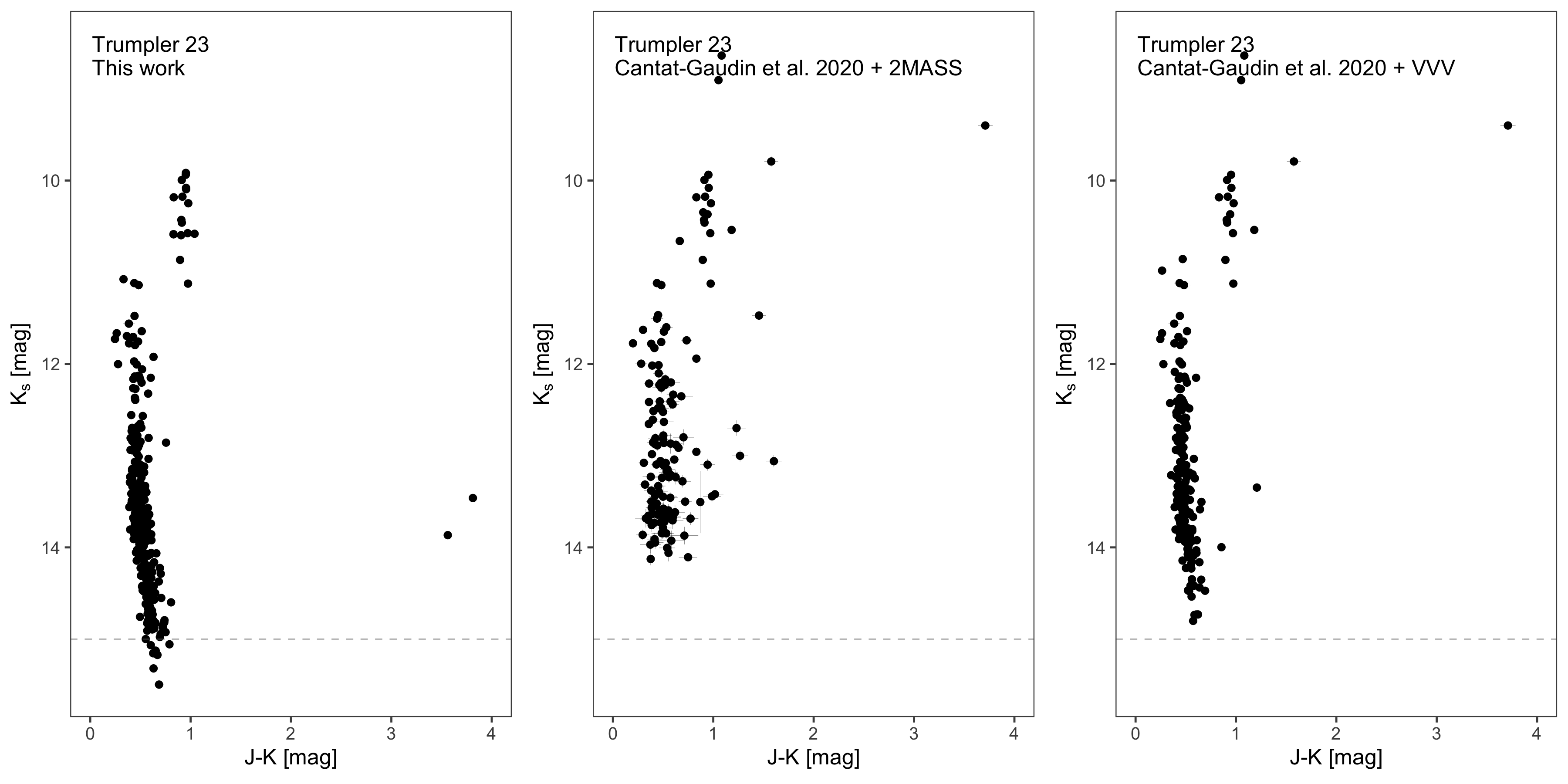}
\caption{Comparison of $K_s$ vs. $J-K_s$ color-magnitude diagrams for NGC\,6067, NGC\,6259, NGC\,4815, Pismis\,18, Trumpler\,23. \textit{Left:} Our VVV data following the procedure outlined in Section\,\ref{gmm_upmask}. \textit{Middle:} Members from \citet{cantatgaudin20} correlated with 2MASS data. \textit{Right}: Members from \citet{cantatgaudin20} directly correlated with our VVV photometry. The horizontal dashed line shows the depth of the cluster sequences at $K_s=15$\,mag. Only members with probabilities $p\geq$90\% are shown.}
\label{fig:all_comp}
\end{figure*}


\end{document}